\documentclass[12pt]{article}

\usepackage{indentfirst}
\usepackage{graphicx}
\usepackage{longtable}
\usepackage{supertabular}
\usepackage{amsmath}
\usepackage{autobreak}
\usepackage{amssymb}
\usepackage{xcolor}
\usepackage{color}
\usepackage{mathrsfs}
\usepackage{amstext}
\usepackage{cases}
\usepackage{float}
\usepackage{threeparttable}
\usepackage{booktabs}
\usepackage[margin=.8in]{geometry}
\usepackage{tikz}
\usetikzlibrary{arrows, graphs}
\usetikzlibrary{calc}
\usetikzlibrary{shapes.geometric, positioning, arrows}
\usepackage[all]{xy}
\usepackage{wrapfig}
\usepackage{subfigure}
\usepackage{caption}
\usepackage{setspace}
\usepackage{multirow}
\usepackage{array}
\usepackage{mathrsfs}

\begin{document}

\title{{\Large\bf {A new social welfare function with a number of desirable properties}}}

\author{Fujun Hou \thanks{Tel.: +86 10
6891 8960; Fax: +86 10 6891 2483; email: houfj@bit.edu.cn.} \\
School of Management and Economics\\
Beijing Institute of Technology\\
Beijing, China, 100081}
\date{}
\maketitle

\begin{abstract}
By relaxing the dominating set in three ways (e.g., from “each member beats every non-member” to “each member beats or ties every non-member, with an additional requirement that at least one member beat every non-member”), we propose a new social welfare function, which satisfies a number of desirable properties including Condorcet winner principle, Condorcet loser principle, strong Gehrlein-stability (hence Smith set principle), anonymity, neutrality, weak Pareto, strong Pareto, non-dictatorship, and [independence of irrelevant alternatives (IIA) when the pairwise majority relation is an ordering on the alternative set]. If the pairwise majority relation is complete and transitive, the proposed method yields a collective preference relation that coincides with the input majority relation. It thus shares the same collective preference function on the dichotomous domain with the approval voting and the majority voting. It runs in polynomial time and thus possesses a competitive advantage over a number of computationally intractable voting rules such as the Dodgson's rule, the Kemeny's rule, the Slater's rule, the Banks rule, and the Schwartz’s tournament equilibrium set (TEQ) rule. When it is used in tournaments, its winner belongs to the uncovered set, the top cycle set, the Smith set, and the Schwartz set. In addition, in a tournament where the number of alternatives is not more than 4, its winner set is a subset, sometimes proper, of the Copeland winner set. Whether this attractive argument is still valid in four-more-alternative tournaments remains an open question.

{\em Keywords}: Social welfare function, dominating set, Arrow's impossibility theorem, uncovered set, tournament
\end{abstract}


\setlength{\baselineskip}{1.5em} 

\section{Introduction}

Ever since 1950s, Arrow's impossibility theorem has become the predominant influence on the development of the modern social choice theory. Arrow's impossibility theorem (Arrow, 1951, 1963) states that, if there are at least three alternatives and the domain of individual preference orderings is unrestricted (U), then there is no transitive-valued social welfare function (SWF, i.e., a mapping from individual orderings to social orderings) satisfying the conditions of weak Pareto (WP), independence of irrelevant alternatives (IIA), and non-dictatorship (ND).

Though no SWF is by any means ``perfect'', it is possible to identify some appealing SWFs with desirable properties. When applied to voting, a SWF is usually called a voting rule. Customarily, the voting rules are divided into two groups, of which the former is of positional rules, and the latter is of binary rules (see, e.g., Fishburn, 1971; Fishburn and Gehrlein, 1976; Favardin et al., 2002; P$\acute{e}$rez-Fern$\acute{a}$ndez and De Baets, 2018; Nurmi, 2023). Roughly speaking, methods of the first group determine the winners by assigning scores to alternatives according to their positions in the individuals' preference rankings, whereas the second group determines winners by means of pairwise comparisons (usually under the simple majority principle). They both have virtues and drawbacks (Nurmi, 1999, 2012, 2023; Gehrlein and Lepelley, 2010; Gaertner, 2017). For example, as a prominent representative of positional rule, the Borda method has many appealing features but may not pick a Condorcet winner (CW) when it exists. Similarly, binary rules, also called Condorcet extensions, are usually able to select the CW as the winner when it exists. However, a Condorcet extension cannot avoid the no-show paradox (Moulin, 1988).

This paper is concerned with constructing a new SWF and disclosing its features. Our main objective is NOT trying to reconcile the discrepancy between the above mentioned two groups. Rather, we will demonstrate how to design a new Condorcet extension with a number of desirable properties.  By the term `desirable' \textbf{we do not mean that the method is fully advisable} since, as previously mentioned, every voting method has virtues and disadvantages of its own.

Recall that a CW is an alternative that {\bf beats every other} alternatives under the majority principle, and that if a CW exists then it is unique (Condorcet, 1785). This concept of CW had ever been extended by researchers from a single winner to a dominating set: A \textbf{dominating set} is defined as a nonempty subset of alternatives all of whose members are preferred to every non-member by some majority of the voters\footnote{The Condorcet winner (CW) concept had ever been extended from a single winner to a winner set, such as the \textit{Condorcet set by Good (1971)}, the \textit{top cycle by Schwartz(1972)}, the Smith set by Smith (1973); the \textit{Condorcet committee by Fishburn (1981)}, and the \textit{Condorcet committee by Gehrlein (1985)}, etc. The one due to Gehrlein (1985) is also called CCG (Condorcet Committee $\grave{a} ~la$ Gehrlein) in the literature (Diss et al., 2020).}. A dominating set (if it exists) can be a good reference for social choice (Diss et al., 2020; Dutta, 1988; Fishburn, 1977; Smith, 1973).

We begin with the dominating set concept and make an attempt to relax it in several novel ways\footnote{The beat-or-tie relaxation of the dominating set is also known as the Kelly set (Kelly, 1977) or the Schwartz set (or GOCHA, see Schwartz (1986)). However, each of our relaxations imposes some additional constraints on the `beat-or-tie relaxation'.}. Particularly, three relaxations will be considered:
\begin{itemize}
\item Extend the bipartition from `each member beats every non-member' to `each member beats or ties every non-member, with an additional requirement that at least one member beat every non-member';
\item Extend the bipartition from `each member beats every non-member' to `each member beats at least one non-member, with an additional requirement that at least one member beat every non-member';
\item Extend the bipartition that `each member beats every non-member' to a tripartition that: on one hand, each member of the top subset beats or ties every member of the middle subset, with an additional requirement that at least one member in the top subset beat every member in the middle subset; on the other hand, each member of the middle subset beat every member of the bottom subset, with an additional requirement that at least one member in the top subset beat every member in the bottom subset.
\end{itemize}

The above relaxations specify some kinds of ordered partitions of alternative set. To avoid exhaustive search for all possible ordered partitions, we further introduce several ``specific'' ordered partitions which are defined with respect to each single alternative. We then connect the specific ordered partitions with alternative weights in an iterative way:
\begin{itemize}
\item [$\diamond$] every alternative may or may not get scores from an alternative-specific ordered partition;
\item [$\diamond$] if the alternative set is not able to be partitioned with respect to a certain alternative, then all alternatives will get zero points from the  ``specific partition'' corresponding to the certain alternative;
\item [$\diamond$] if the alternative set can be partitioned with respect to the certain alternative, then the alternatives in the bottom block (``bottom block'' will be interpreted in Section 3) of the ordered partition will get no points, while the other alternatives will get some positive points.
\end{itemize}
The alternatives' overall scores will be used for ranking and selecting. Our idea will result in a ``scoring method'' which assigns scores to alternatives according to their ``block positions'' in possible ordered partitions rather than ``ranking positions'' in individual preference rankings. It will have a number of attractive properties which will be set out at length.

More specifically, given a complete (connected) majority preference relation (not necessarily transitive) over a finite alternative set, we first define two parameters which can only take values in [0,1]. We show that, when a Condorcet winner or, more generally, a dominating set exists, those defined parameters would take the value 1, which specifies the feature of a dominating set that ``each member beats every non-member''. Second, we introduce some ordered partitions of the alternative set (called \textbf{dominating-set-relaxed partitions, for short, DSRPs}) by only requiring the parameter values to be higher than 0, which will reflect the idea of our relaxations. For instance, a positive value of a parameter will indicate that ``at least one alternative in one set beats every alternative in another set''. Obviously, the amount of computation required by searching for all possible partitions grows exponentially with the number of alternatives. In order to avoid such troubles, we will introduce ``specific'' partitions which are defined with respect to a single alternative. Since a majority preference relation may correspond to none, a unique or even multiple specific DERPs, we will propose a scoring method (called \textbf{the dominating-set-relaxed-scoring method, for short, DSR-scoring method}) for evaluating alternatives by using the specific DSRPs. It will be proved that the proposed DSR-scoring method runs in polynomial time of $O(m^3)$ with $m$ being the number of alternatives. Third, we will design a ranking and choosing procedure which first (in its first sub-procedure) transforms the individual binary relations into a pairwise majority relation and then (in its second sub-procedure) obtains the overall evaluation of the alternatives by using the DSR-scoring method. In this scenario, since the procedure is transitive-valued, it can be used as a SWF or a SCC (social choice correspondence, i.e., a mapping from individual orderings to social choice sets). When the DSR-scoring method is used as a SWF/SCC, we will prove that it satisfies a number of desirable properties such as Condorcet winner principle, Condorcet loser principle, strong Gehrlein-stability (hence Smith set principle), anonymity, neutrality, weak Pareto, strong Pareto, and non-dictatorship. Given the individual preference orderings over a finite universal set, if the pairwise majority relation is also an ordering over the same universal set, then the SWF/SCC will satisfy the condition of IIA. The DSR-scoring method also suits for the situation, as a particular case, where the individual preferences are restricted to be dichotomous as can be easily inferred from approval ballots (Brandl and Peters 2022). In addition, the DSR-scoring method can be used in a setting of tournament. The comparison with some existing methods (such as the Copeland method, the Dodgson's rule, and the Kemeny's rule, etc.) and the relationships with some well-studied choice set concepts (such as the uncovered set, the top cycle set, and the Smith set, etc.) will be investigated.

The paper is structured as follows. Some preliminary and notations will be set out in Section 2. The above mentioned ordered partitions and specific ordered partitions will be formally defined in Section 3. An algorithm for seeking specific ordered partitions is outlined which is proved to run in polynomial time. Section 4 includes the strategy for how to connect alternatives' scores with alternative-specific partitions, the formula for computing overall scores of alternatives, and the analysis of computational complexity. Section 5 contains possible application scenarios of the proposed method and the detailed proofs of desired properties. Section 6 performs a comparison with some existing methods, and analyses the relationships between the proposed method and some well-studied choice sets. Section 7 contains our concluding remarks. Numerous illustrative examples are presented throughout the paper.

\section{Notation and preliminary}

To conduct the discussion, the following \textit{notations} will be used, most of which are contained in Campbell and Kelly (2002), Barber$\grave{a}$ (2011) and Patty and Penn (2019).
\begin{itemize}
\item [-] $X=\{x_1,x_2,\ldots,x_m\}$ stands for the \textbf{alternative} (candidate, outcome, protocol, etc.) set with $1<m<+\infty$. Let $2^{X}_+$ represent the power set of $X$ with the empty set excluded. Sometimes we also use $x,y,z,u,w,\ldots$ to stand for alternatives, and $X_t,X_b,Y,A,B,C,\ldots$ for alternative subsets in $2^{X}_+$. If $A$ is a subset of $X$, we let $|A|$ stand for the cardinality of $A$, and $X\setminus A$ the complement of $A$ in $X$.
\item [-] $N=\{1,2,\ldots,n\}$ stands for the \textbf{individual (voter)} set, i.e., the \textbf{society}, with $1<n<+\infty$.
\item [-] We will use symbols `$\land$', `$\lor$' and `$\neg$' for the logical operators AND, OR and NOT, respectively.
\item [-] Let $R(X)$ denote a \textbf{binary relation} on $X$, which is interpreted as a \textbf{preference relation} in the sense that $x\succ y$, $x\sim y$ and $x\prec y$ represent ``$x$ beats $y$ (i.e., $x$ dominates $y$ in the sense that $x$ is preferred to $y$)", ``$x$ ties $y$ (i.e., $x$ is indifferent to $y$)" and ``$x$ is beaten by $y$ (i.e., $x$ is less preferred to $y$)", respectively. When necessary we will use subscript `$k$' to indicate individual $k$'s preferences  such as $\succ_k$, $\sim_k$ and $\prec_k$. When it is for the social preference, similar notations are used without  any subscript. For simplicity, the symbols  `$\succ$' and `$\sim$' may refer to a preference of an individual, the preference of a society, or even a preference or a pairwise comparison outcome in a very general sense, depending on the context in which they are used. 
\item [-] A binary relation $R(X)$ is \textbf{complete} if for all $x,y\in X$ with $x\neq y$ there exists exactly one of the three preferences $x\succ y$, $x\sim y$ and $x\prec y$. $R(X)$ is \textbf{transitive}\footnote{In addition to P39 of Campbell and Kelly (2002), the reader may also refer to Arrow (1963, P14).} if for all $x,y,z\in X$:
\begin{itemize}
\item [] $(x\succ y \land y\succ z)\rightarrow x\succ z$.
\item [] $(x\sim y \land y\succ z)\rightarrow x\succ z$.
\item [] $(x\succ y \land y\sim z)\rightarrow x\succ z$.
\item [] $(x\sim y \land y\sim z)\rightarrow x\sim z$.
\end{itemize}
If $R(X)$ is complete and transitive then it is called an \textbf{ordering} of the alternatives and will be denoted by $O(X)$. A strict (i.e., linear) ordering is denoted by $L(X)$. Let $\mathcal{R}(X)$, $\mathcal{O}(X)$, and $\mathcal{L}(X)$ denote the set of all complete binary relations on $X$, the set of all weak orderings on $X$, and the set of all strict orderings on $X$, respectively. A preference \textbf{profile} is a map $\varrho(X): N\rightarrow \mathcal{O}(X)$, which assigns an ordering to each individual. Thus a profile is an element of $\mathcal{O}^n(X)$ that can be written as a list $(O_1(X),O_2(X),\ldots,O_n(X))$. A non-empty subset $\mathcal{D}(X)$ of $\mathcal{O}^n(X)$ is called a \textbf{domain}. In order to indicate individual $k$'s preference over $\{x,y\}$ at profile $\varrho(X)$, we will write, for instance, $x\succ^{\varrho(X)}_{k} y$.
\item [-] A \textbf{preference aggregation rule} $f$ assigns a complete binary relation on the alternative set to each preference profile, i.e., $f: \mathcal{D}(X)\rightarrow \mathcal{R}(X)$. So $f(\varrho(X))$ describes a social preference relation on $X$ at profile $\varrho(X)$. If $f(\varrho)$ is an ordering on $X$, then $f$ becomes an Arrovian \textbf{social welfare function (SWF)}, i.e., $f: \mathcal{D}(X)\rightarrow \mathcal{O}(X)$. If either $\mathcal{D}(X)=\mathcal{O}^n(X)$ or $\mathcal{D}(X)=\mathcal{L}^n(X)$, then we say that $f$ has a full domain. Otherwise, $\mathcal{D}(X)$ will be a restricted domain. A \textbf{social choice function (SCF)} on the domain $\mathcal{D}(X)\subseteq\mathcal{O}^n(X)$ is a function\footnote{There are several different ways to define a SCF in literature, for example, one may refer to Arrow (1963, P15).} $g: \mathcal{D}(X)\rightarrow X$, which assigns one alternative as the optimal outcome to each admissible profile in the domain $D$ (Barber$\grave{a}$, 2011, P739). A \textbf{social choice correspondence (SCC)} on the domain $\mathcal{D}(X)\subseteq\mathcal{O}^n(X)$ is a function $h: \mathcal{D}(X)\rightarrow 2^{X}_+$, which assigns a non-empty set of alternatives as the optimal outcome (also known as the solutions set or the choice set) to each admissible profile in the domain $D(X)$ (Barber$\grave{a}$, 2011, P788). If it is endowed with a tie-breaking rule, a SWF or a SCC can work as a SCF.
\item [-] Suppose $Y$ is a non-empty subset of $X$, i.e., $Y\in 2^X_+$. Let $\varrho(X)|Y$ denote the \textbf{restriction of profile} $\varrho(X)$ to subset $Y$, which indicates the individuals' preferences only concerning the alternatives in $Y$. Let $f(\varrho(X))|Y$ denote the \textbf{restriction of social preference relation} $f(\varrho(X))$ to subset $Y$, which indicates the social preferences only concerning the alternatives in $Y$. Given a profile $\varrho(X)$ and a preference aggregation rule $f$, the social preference over $\{x,y\}\subset X$ can be described by $f(\varrho(X))|\{x,y\}$ or $x f(\varrho(X)) y$. For instance, if $f(\varrho(X))$ ranks $x$ strictly higher than $y$, then we will write $x\succ_{f(\varrho(X))} y$.
\item [-] For two disjoint nonempty alternative subsets $A$ and $B$, by $A\succ B$ we mean that each alternative in $A$ beats each alternative in $B$ (Good, 1971; Schwartz, 1972; Smith, 1973; Gehrlein, 1985). Similar interpretation is attached to $A\succeq B$ (Kelly, 1977; Schwartz, 1986).
\item [-] $N(x_i\succ_k x_j)$ and $N(x_i\sim_k x_j)$ denote the numbers of individuals whose preferences are $x_i\succ_k x_j$ and $x_i\sim_k x_j$, respectively. For the sake of convenience, the pairwise comparison results of $N(x_i\succ_k x_j)$ are usually tallied in a matrix called the \textbf{pairwise comparison matrix}\footnote{In literature, it is also called voting matrix (see, e.g., Young, 1988) or outranking matrix (see, e.g., Arrow and Raynaud, 1986).} (see, e.g., Young, 1995; Nurmi, 1999, P7).
\end{itemize}

In addition, the following {\textit{preliminaries}} will be involved.
\begin{itemize}
\item [-] The \textbf{majority rule}\footnote{The majority rule in this paper belongs to a `simple majority rule' rather than an `absolute majority rule'. For knowledge of the difference, readers may refer to Gaertner (2009, P39).} $M(X)$ means that $x_i\succ_M x_j$ iff $N(x_i\succ_k x_j)>N(x_j\succ_k x_i)$, and $x_i\sim_M x_j$ iff $N(x_i\succ_k x_j)=N(x_j\succ_k x_i)$.
\item [-] A \textbf{cycle} refers to a dominating relation such as $x_p\succ x_q\succ, \ldots, \succ x_r\succ x_p$. A nonempty subset $B$ of alternative set $X$ is defined as a \textbf{top cycle set} such that $B$ is a singleton (which contains the Condorcet winner) or a dominating cycle relation (i.e., $B=\{x_1,x_2,\ldots,x_t\}$, and $x_1\succ x_2\succ\ldots\succ x_t\succ x_1$ for some $t$ and $x_1,x_2,\ldots,x_t$), and $x\succ y$ for all $x\in B$  and $y\in (X\setminus B)$ (Schwartz, 1972,1990).
\item [-] A tournament is a complete binary relation on alternative set $X=\{x_1,x_2,\ldots,x_m\}$ without ties (Rubinstein, 1980). A \textbf{tournament matrix}\footnote{A tournament matrix can also be defined as a $m\times m$ 0-1 matrix whose [i,j] entry has the value of 1 whenever $x_i\succ x_j$ and 0 otherwise (see, e.g., Nurmi (1999, P7)).} $T(X)$ is defined as a $m\times m$ matrix whose [i,j] entry has the value of 1 whenever $x_i\succ x_j$, $-1$ whenever $x_i\prec x_j$, and 0 whenever $i=j$ (see, e.g., Dutta, 1988).
\end{itemize}

\begin{itemize}
\item [-] \textbf{Weak Pareto principle (WP)}: if every individual prefers alternative $x$ to alternative $y$, then so does the society. Formally, $\forall\varrho(X)\in\mathcal{D}(X)\forall (x,y)\in X\times X[(x\succ_k^{\varrho(X)} y, \forall k\in N)\rightarrow(x\succ_{f(\varrho(X))} y)]$. 
\item [-] \textbf{Strong Pareto principle (SP)}: if `$x$ is preferred or indifferent to $y$' holds for all individuals and `$x$ is preferred to $y$' holds for some individual, then $x$ is socially preferred to $y$. Formally, $\forall\varrho(X)\in\mathcal{D}(X)\forall (x,y)\in X\times X\{[(x\succeq_k^{\varrho(X)} y, \forall k\in N)\land(x\succ_k^{\varrho(X)} y, \exists k\in N)]\rightarrow(x\succ_{f(\varrho(X))} y)\}$.
\item [-] \textbf{Non-dictatorship (ND)}: there does not exist such an individual that if she prefers $x$ to $y$ then the society must prefer $x$ to $y$. Formally, $\neg\exists k\{(k\in N)\land[\forall\varrho(X)\in\mathcal{D}(X)\forall (x,y)\in X\times X(x\succ_k^{\varrho(X)} y\rightarrow x\succ_{f(\varrho(X))} y)]\}$.
\item [-] \textbf{Independence of irrelevant alternatives (IIA)}: the social preference of two alternatives depends only upon the individuals' preferences concerning those two alternatives, Formally, $\forall (\varrho(X),\varrho'(X))\in\mathcal{D}(X)\times\mathcal{D}(X)\forall (x,y)\in X\times X[\varrho(X)|\{x,y\}=\varrho'(X)|\{x,y\}\rightarrow f(\varrho(X))|\{x,y\}=f(\varrho'(X))|\{x,y\}]$.
\item [-] \textbf{Condorcet winner principle}:  when a Condorcet winner (CW) exists \footnote{As shown by Condorcet himself, the Condorcet winner may not always exist (see, Condorcet, 1785, or, Young, 1995)}, it is unique, and Condorcet suggests that it be the choice when it exists. This is the well-known Condorcet winner principle. The Condorcet winner refers to an alternative $x$ that can beat every other alternative by pairwise majority. Formally,
$$\forall y\in X\setminus\{x\}: x\succ_M y.\eqno(1)$$
\item [-] \textbf{Condorcet loser principle}\footnote{Borda suggested that a CL not be selected. The phenomenon that an election method may not guarantee the exclusion of a CL is called the Borda paradox (Nurmi, 1999, P13).}: when the Condorcet loser exists, it should never be elected. The Condorcet loser (CL) refers to an alternative that is beaten by every other alternative by pairwise majority. 
\item [-] \textbf{Strongly Gehrlein-stability}:  A voting rule is strongly Gehrlein-stable if it always elects a Condorcet committee $\grave{a} ~la$ Gehrlein (dominating set) whenever such a set exists (Barber$\grave{a}$ and Coelho, 2008). Gehrlein (1985) extends the concept of Condorcet winner to Condorcet committee: each member in the committee beats each non-member by majority rule. A Condorcet committee $\grave{a} ~la$ Gehrlein is also known as a \textbf{dominating set} (Brandt, 2009). Formally, a non-empty alternative set $A$ is called a dominating set if
$$A\succ X\setminus A, \mbox{~namely}, \forall x\in A, \forall y\in X\setminus A: x\succ_M y.\eqno(2)$$
\item [-] \textbf{Smith set principle}:  A voting rule is said to satisfy the Smith set principle in the sense that it only selects those alternatives that are included in all dominating sets (Smith, 1973; Fishburn, 1977; Brandt, 2009). The Smith set refers to the smallest set where each member beats every non-member (Brandt et al., 2007). Obviously, strong Gehrlein-stability implies Smith set principle.

\end{itemize}

\section{Dominating-set-relaxed partitions (DSRPs)}

\subsection{Motivation and two parameters}

Our \textbf{motivation} comes from the following observation and analysis on the formulations of Condorcet winner and dominating set (namely, equations (1) and (2)). Assume that we have a complete majority relation $M(X)$ on $X$. Let $A$ and $B$ be nonempty disjoint subsets of alternative set $X$ (formally, $A, B\in 2^X_+$, and $A\cap B=\emptyset$). 
\begin{itemize}
\item [(1)] If we define \textbf{two parameters}
$$\beta(A,B)=\frac{\big |\{x\in A | \forall y\in B, x\succ_M y \}\big |}{\big | A\big |},\eqno(3)$$
and
$$\gamma(A,B)=\min_{x\in A}\frac{\big |\{y\in B | x\succ_M y \}\big |}{\big | B\big |}, \eqno(4)$$
then we know
\begin{itemize}
\item  $\beta(A,B)\in [0,1]$. In particular,
\begin{itemize}
\item $\beta(A,B)>0$ when there exists at least one alternative in $A$ that beats all the alternatives in $B$.
\item $\beta(A,B)=1$ when each alternative in $A$ beats all alternatives in $B$. Thus in this case (i.e., $\beta(A,B)=1$) if $A=X\setminus B$, then $A$ is a dominating set.
\end{itemize}
\item  $\gamma(A,B)\in [0,1]$. In particular,
\begin{itemize}
\item $\gamma(A,B)>0$ when each alternative in $A$ beats at least one alternative in $B$.
\item $\gamma(A,B)=1$ when each alternative in $A$ beats all alternatives in $B$. Thus in this case (i.e., $\gamma(A,B)=1$) if $A=X\setminus B$, then $A$ is a dominating set.
\end{itemize}
\item  $x$ is a CW iff (i) $x$ is a singleton of $A$; (ii) $B=X\setminus A$; and (iii) $\beta(A,B)=1$.
\item  $A$ is a dominating set iff (i) $B=X\setminus A$; and (ii) $\beta(A,B)=1$.
\end{itemize}
One can see that $\beta(A,B)=1$ and $\gamma(A,B)=1$ have identical meanings, thus we also have
\begin{itemize}
\item  $x$ is a CW iff (i) $x$ is a singleton of $A$; (ii) $B=X\setminus A$; and (iii) $\gamma(A,B)=1$.
\item  $A$ is a dominating set iff (i) $B=X\setminus A$ and (ii) $\gamma(A,B)=1$.
\end{itemize}
\item [(2)] The above analysis indicates that a dominating set which is described by Eq.(2) implies $A\succ B$, $\beta(A,B)=1$, and $\gamma(A,B)=1$. Therefore, \textbf{three intuitive ways to extend the concept of dominating set} might be considered as follows:
\begin{itemize}
\item  One way is to merely require ``$A\succeq B$'' and $\beta(A,B)>0$ by which we can define \textbf{an ordered bipartition} $\left \langle A,B\right \rangle$, where $\beta(A,B)>0$ and $A\succeq B$\footnote{When $B=X\setminus A$, the expression $A\succeq B$ is also referred to as the Kelly set (Kelly, 1977),  the Schwartz set (or GOCHA, see Schwartz (1986)), the ``Condorcet set'' (Miller, 1977), the ``weak Condorcet set $\grave{a} ~la$ Gehrlein'' or ``weak Condorcet committee $\grave{a} ~la$ Gehrlein'' (see, e.g., Barber$\grave{a}$ and Coelho, 2008; Coelho, 2005; Kamwa, 2017; Ratliff, 2003).}. By $\beta(A,B)>0$ we mean ``all alternatives in $B$ are defeated by at least one alternative in $A$''. By the bipartition `$\left \langle A,B\right \rangle$, where $\beta(A,B)>0$ and $A\succeq B$' we extend the dominating set concept from ``all alternatives in $A$ are preferred to all alternatives in $X\setminus A$'' to ``all alternatives in $A$ are preferred or indifferent to all alternatives in $X\setminus A$; moreover, all alternatives in $X\setminus A$ are defeated by at least one alternative in $A$''.
\item  A second way is to merely require $\beta(A,B)>0$ and $\gamma(A,B)>0$ by which we can define \textbf{another form of ordered bipartition} $\left \langle A,B\right \rangle$, where $\beta(A,B)>0$ and $\gamma(A,B)>0$. By $\gamma(A,B)>0$ we mean ``each alternative in $A$ defeats at least one alternative in $B$''. By the bipartition `$\left \langle A,B\right \rangle$, where $\beta(A,B)>0$ and $\gamma(A,B)>0$' we extend the dominating set concept from ``all alternatives in $A$ are preferred to all alternatives in $X\setminus A$'' to ``all alternatives in $X\setminus A$ are defeated by at least one alternative in $A$''; moreover, ``each alternative in $A$ defeats at least one alternative in $X\setminus A$''.
\item We can have a third way as follows. Suppose $B$ can be partitioned into two subsets $B_1$ and $B_2$ such that $B_1\succ B_2$. Accordingly, given Eq.(2) we will have $A\succ (B_1\cup B_2)$, $\beta(A, B_1\cup B_2)=1$ and $B_1\succ B_2$. We further relax $A\succ (B_1\cup B_2)$ and $\beta(A, B_1\cup B_2)=1$ by replacing them by $A\succeq B_1$ and $\beta(A, B_1\cup B_2)>0$, respectively. We go one step further to replace $\beta(A, B_1\cup B_2)>0$ by $\beta(A, B_1)>0$ and $\beta(A, B_2)>0$. By doing so we will define \textbf{an ordered tripartition} $\left \langle A,B_1,B_2\right \rangle$ where $A\succeq B_1$, $\beta(A, B_1)>0$, $B_1\succ B_2$, and $\beta(A,B_2)>0$\footnote{In a tournament (connected and no ties), $A\succeq B_1$ implies $A\succ B_1$. Then [$A\succeq B_1$, $\beta(A, B_1)>0$, $B_1\succ B_2$, and $\beta(A,B_2)>0$] means that, in Miller's terms (Miller, 1980), there exists at least one alternative in $A$ such that it \textit{covers} every alternative in $B_1$.}. This sort of ordered tripartition allows a desirable flexibility for dealing with cycles whose elements may belong to separate blocks since a tripartition $\left \langle A,B_1,B_2\right \rangle$ may allow such a cycle as $x\succ_M y$, $y\succ_M z$ and $z\succ_M x$, where $x\in A$, $y\in B_1$, $z\in B_2$ and $|A\cup B_1\cup B_2|>3$.
\end{itemize}
\end{itemize}

The above extensions are rather general since the values of $\beta$ and $\gamma$ are no longer limited to 1 (as the case of a dominating set). In what follows, the above mentioned three types of ordered partition will be formally defined and will be called the \textit{dominating-set-relaxed partitions (for short, DSRPs)} since they originate from the concept of dominating set.

\subsection{Dominating-set-relaxed bipartition of two types}

Suppose that the complete pairwise comparison outcomes between the alternatives are provided (e.g., as indicated by Henriet (1985), a voter's preference, social preferences by using the majority rule, tournaments, or some other sort of complete outcomes in pairwise contests). In what follows we may omit the subscript `$M$' so that the preference relation can refer to whatever outcomes resulted from pairwise comparisons. Now we formally define three sorts of ordered partitions alluded to in previous subsection.

\textbf{Definition 1} Provided that a complete binary preference relation $R(X)$ on alternative set $X$ where $1<\lvert X \rvert <+\infty$ is given. An ordered set pair $\left \langle A, B\right \rangle$ is called a \textbf{first-type dominating-set-relaxed bipartition} of $X$ such that
\begin{itemize}
\item [(1)] $A\in 2^X_+$, $B\in 2^X_+$, $A\cap B=\emptyset$, $A\cup B=X$;
\item [(2)] $\beta(A,B)>0$, that is, $\exists x\in A(\forall y\in B\rightarrow x\succ y)$;
\item [(3)] $A\succeq B$, that is, $\forall x\in A$, $\forall y\in B$: $x\succeq y$.
\end{itemize}

\textbf{Definition 2} Provided that a complete binary preference relation $R(X)$ on alternative set $X$ where $1<\lvert X \rvert <+\infty$ is given. An ordered set pair $\left \langle A, B\right \rangle$ is called a \textbf{second-type dominating-set-relaxed bipartition} of $X$ such that
\begin{itemize}
\item [(1)] $A\in 2^X_+$, $B\in 2^X_+$, $A\cap B=\emptyset$, $A\cup B=X$;
\item [(2)] $\beta(A,B)>0$, that is, $\exists x\in A(\forall y\in B\rightarrow x\succ y)$;
\item [(3)] $\gamma(A,B)>0$, that is, $\forall x(x\in A \rightarrow (\exists y\in B, x\succ y))$.
\end{itemize}

\vspace{0.2cm}
We refer to both the first- and second-type dominating-set-relaxed bipartitions as \textbf{dominating-set-relaxed bipartitions}\footnote{Elkind et al. (2015) proposed a set-valued relaxation of Condorcet winner called \textit{Condorcet winning set}. It is defined as set $Y$ such that ``\textit{if for every alternative $z$ in $X\setminus Y$, a majority of voters prefer some alternative in $Y$ to $z$}''(Elkind et al., 2015). The concept of dominating-set-relaxed bipartition differs from the Condorcet winning set in that, the first block $Y$ of a dominating-set-relaxed bipartition $\langle Y,X\setminus Y\rangle$ is always a Condorcet winning set, but a Condorcet winning set and its complement do not necessarily constitute of a dominating-set-relaxed bipartition.}. To illustrate, we consider the following example.

\textbf{Example 1} Suppose we have a complete binary relation on $X=\{a,b,c,d\}$ as shown by Figure 1, where `$x\rightarrow y$' indicates $x\succ y$, and `$x\dashrightarrow y$' indicates $x\sim y$. According to Def.1 and Def.2, we know that $\left \langle \{a,b,c\},\{d\}\right \rangle$, $\left \langle \{a,b\},\{c,d\}\right \rangle$, and $\left \langle \{a\},\{b,c,d\}\right \rangle$ are all dominating-set-relaxed bipartitions of $\{a,b,c,d\}$. In particular, $\left \langle \{a,b,c\},\{d\}\right \rangle$ belongs to the first-type, $\left \langle \{a,b\},\{c,d\}\right \rangle$ belongs to the second-type, and $\left \langle \{a\},\{b,c,d\}\right \rangle$ belongs to both types of bipartitions.

\begin{figure}[H]
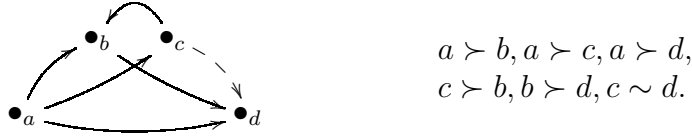

\centering
\subfigure{}
{
\begin{minipage}{5cm}
\[ \xygraph{  
        !{<0cm,0cm>;<1cm,0cm>:<0cm,1cm>::}  
        !{(0,0) }*+{\bullet_{a}}="a"  
        !{(1,1) }*+{\bullet_{b}}="b"  
        !{(2,1) }*+{\bullet_{c}}="c"  
        !{(3,0)}*+{\bullet_{d}}="d"  
            "a":@/^0.2cm/"b"^(0.6){}
            "a":@/_0.1cm/"c"^(0.6){}
            "b":@/_0.1cm/"d" ^(0.8){}   
          "c":@/^0.2cm/@{-->}"d" ^(0.8){}  
            "a":@/_/"d" ^(0.4){}  
            "c":@/_0.5cm/"b" ^(0.7){}   
        }
\]  
\end{minipage}
}
\subfigure{}
{
\begin{minipage}{5cm}
\[
\begin{array}{ll}
 & a \succ b, a\succ c, a\succ d,\\
 & c \succ b, b\succ d, c\sim d.
\end{array}
\]
\end{minipage}
}
\caption{Alternative-relationship diagram of Example 1.}
\end{figure}

\vspace{0.2cm}
Since a Condorcet winner (CW) beats all other alternatives in a pairwise majority comparison as shown by expression (1), from Def.1 and Def.2 we have the following result.

\textbf{Proposition 1} Alternative $x$ is a CW iff $\left \langle \{x\},X\setminus\{x\}\right \rangle$ is a dominating-set-relaxed bipartition of alternative set $X$ under a complete majority relation $M(X)$.

\textbf{Proof} The proof we offer here is for the first-type dominating-set-relaxed bipartition. The proof is similar when it is for the second-type. On the one hand, if $\{x\}$ and $X\setminus\{x\}$ constitute a dominating-set-relaxed bipartition of $X$ then, from Def.1, we always have $\beta(\{x\},X\setminus\{x\})>0$. Resultantly, we have $\{x\}\succ X\setminus\{x\}$ since $\beta(A,B)>0$ implies that there exists at least one alternative in $A$ that beats all alternatives in $B$. Hence we know that $x$ is a CW.

On the other hand, if $x$ is a CW then: (a) we know that $\{x\}$ and $X\setminus\{x\}$ constitute a partition of $X$ and thus the first item of Def.1 is fulfilled; and (b) we have $\{x\}\succ X\setminus\{x\}$ hence $\beta(\{x\},X\setminus\{x\})>0$ and $\{x\}\succeq X\setminus\{x\}$. By Def.1 we know that $\left \langle \{x\},X\setminus\{x\}\right \rangle$ is a dominating-set-relaxed bipartition. Q.E.D.

\vspace{0.2cm}
The above proposition asserts the fact that the dominating-set-relaxed bipartition is an extension of CW. Now that a dominating set is an extension of CW, what is the connection between a dominating set and the dominating-set-relaxed bipartition? The following proposition offers an answer.

\textbf{Proposition 2} Suppose that $A$ is a proper subset of $X$. If $A$ is a dominating set then $\left \langle A, X\setminus A\right \rangle$ is a dominating-set-relaxed bipartition under a complete majority relation.

\textbf{Proof} The proof we offer here is for the first-type dominating-set-relaxed bipartition. The proof is similar when it is for the second-type. Given that $A$ is a proper subset of $X$, we know that $X$ can be partitioned into $A$ and $X\setminus A$ and thus the first item of Def.1 is satisfied by $\left \langle A, X\setminus A\right \rangle$. On the other hand, if $A$ is a dominating set then from expression (2) we have 
$$\forall x\in A, \forall y\in X\setminus A: x\succ y,$$
which implies that: (a) the restriction item (2) of Def.1 is satisfied by $\left \langle A, X\setminus A\right \rangle$ since in this case we have $\beta(A,X\setminus A)=1$; and (b) the restriction item (3) of Def.1 is satisfied by $\left \langle A, X\setminus A\right \rangle$ since in this case we have $A\succ X\setminus A$. Therefore, the proposition holds. Q.E.D.

\vspace{0.2cm}
The previous two propositions assert that the dominating-set-relaxed bipartition is not only an extension of CW but also an extension of the dominating set. 

\subsection{Dominating-set-relaxed tripartition}

\textbf{Definition 3}  Given a complete binary preference relation $R(X)$ on alternative set $X$ where $1<\lvert X \rvert <+\infty$, an ordered set triple $\left \langle A, B, C\right \rangle$ is called a \textbf{dominating-set-relaxed tripartition} of $X$ such that
\begin{itemize}
\item [(1)] $A\in 2^X_+$, $B\in 2^X_+$, $C\in 2^X_+$, $A\cap B=\emptyset$, $A\cap C=\emptyset$, $B\cap C=\emptyset$, $A\cup B\cup C=X$;
\item [(2)] $A\succeq B$, $B\succ C$;
\item [(3)] $\beta(A,B)>0$ and $\beta(A,C)>0$.
\end{itemize}

In Example 1, the following set triples are permissible dominating-set-relaxed tripartitions (on the basis of the majority relations therein):
$$\left \langle \{a,c\},\{b\},\{d\}\right \rangle, \left \langle \{a,d\},\{c\},\{b\}\right \rangle.$$

From Def.1 and Def.3, we have the following result.

\textbf{Proposition 3} If a dominating-set-relaxed tripartition $\left \langle A, B, C\right \rangle$ satisfies $A\succeq C$, then the dominating-set-relaxed tripartition can reduce to a first-type dominating-set-relaxed bipartition $\left \langle A\cup B, C\right \rangle$.

\textbf{Proof} First, the restriction item (1) of Def.1 holds trivially for $\left \langle A\cup B, C\right \rangle$ since $\left \langle A, B, C\right \rangle$ is a tripartition satisfying the restriction item (1) of Def.3. Second, because $\left \langle A, B, C\right \rangle$ is a tripartition we thus have $A\succeq B$ and $B\succ C$. If we also have $A\succeq C$, then we know $(A\cup B)\succeq C$ and thus the restriction item (3) in Def.1 is fulfilled by $\left \langle A\cup B, C\right \rangle$. Finally, since $\left \langle A, B, C\right \rangle$ is a tripartition, we know that $B\neq\emptyset$ and $B\succ C$. Thus we have $\beta(B, C)>0$, and accordingly, $\beta(A\cup B, C)>0$. Hence the restriction item (2) of Def.1 is fulfilled by $\left \langle A\cup B, C\right \rangle$. Therefore, by Def.1 we know that $\left \langle A\cup B, C\right \rangle$ is a bipartition. Q.E.D.

\vspace{0.2cm}

The Example 1 can be a good illustration for Proposition 3. The dominating-set-relaxed tripartition $\left \langle \{a,c\},\{b\},\{d\}\right \rangle$ can, according to Proposition 3, reduce to the dominating-set-relaxed bipartition $\left \langle \{a,b,c\},\{d\}\right \rangle$, since we have $\{a,c\}\succeq \{d\}$ in Example 1.

\vspace{0.2cm}
The proposition 3 asserts the fact that the dominating-set-relaxed tripartition (as defined by Def.3) is an extension of the dominating-set-relaxed bipartition (as defined by Def.1). Because the dominating-set-relaxed bipartition is an extension of CW, thus the dominating-set-relaxed tripartition can also be viewed as an extension of CW.

For convenience of discussion, sometimes we write a dominating-set-relaxed bipartition and a dominating-set-relaxed tripartition in the following forms
$$
\left\langle
\begin{array}{c}
A\\
B
\end{array}
\right\rangle{\mbox{~and~}}
\left\langle
\begin{array}{c}
A\\
B\\
C
\end{array}
\right\rangle,
$$
where we call $A$ the \textit{top block}, $C$ the \textit{bottom block}, and $B$ the \textit{bottom block} in the case of dominating-set-relaxed bipartition and the \textit{middle block} in the case of dominating-set-relaxed tripartition. We refer to both the dominating-set-relaxed bipartition and the dominating-set-relaxed tripartition as the \textbf{dominating-set-relaxed partition (DSRP)}. In addition, for succinctness sometimes we take the term ``partition'' directly to refer to a dominating-set-relaxed bipartition or a dominating-set-relaxed tripartition.

The introduction of the dominating-set-relaxed partitions (DSRPs) is meant to make connections between ordered partitions and choice set. One may be aware of the fact that to write out all possible DSRPs of a large alternative set can be exhausted, since checking the relationship between the subsets of $X$ will be needed. As will be analyzed in Section 4, however, \textbf{some particular DSRPs are adequate for our purpose}, and thus we do not need to deal with all possible DSRPs. In the next subsection, we introduce the specific ordered partitions.

\subsection{Alternative-specific DSRP}

\subsubsection{DSRP with respect to a single alternative}

Suppose that $R(X)$ is a complete binary preference relation on alternative set $X$. With regard to alternative $z$, we introduce the following notations\footnote{The first two notations had ever been used by many researchers as in Miller (1977,1980), Banks (1985) and Fishburn (1990) as what those authors called ``$F(z), D(z)$'', ``$P(z), P^{-1}(z)$'' and ``up set, down set'', respectively.}
$$X_z^{\succ}=\{x\in X\mid x\succ z\}, X_z^{\prec}=\{x\in X\mid x\prec z\}, X_z^{\sim,\neq}=\{x\in X\setminus\{z\}\mid x\sim z\}.$$
For the sake of succinctness, we will also use $X_z^{\succeq,\neq}$ to refer to $X_z^{\succ}\cup X_z^{\sim,\neq}$, $X_z^{\succeq}$ to $\{x\in X\setminus\{z\}\mid x\succeq z\}\cup\{z\}$, $X_z^{\preceq}$ to $\{x\in X\setminus\{z\}\mid x\preceq z\}\cup\{z\}$, and $X_z^{\sim}$ to $\{x\in X\setminus\{z\}\mid x\sim z\}\cup\{z\}$.

Taking advantage of Def.1-Def.3 and the above notations, we have the following result.

\textbf{Theorem 1} Suppose that a complete binary preference relation $R(X)$ on alternative set $X$ where $1<\lvert X \rvert <+\infty$ is given. With respect to alternative $z$, we have
\begin{itemize}
\item [(1)] Provided that $X_z^{\prec}=\emptyset$, then the ordered pair $\left \langle X_z^{\succ}\cup X_z^{\sim,\neq}, \{z\}\right \rangle$ is a dominating-set-relaxed bipartition of $X$ iff $X_z^{\succ}\neq\emptyset$.
\item [(2)] Provided that $X_z^{\prec}\neq\emptyset$, then the ordered pair $\left \langle X_z^{\succeq,\neq}\cup\{z\},X_z^{\prec} \right \rangle$ is a dominating-set-relaxed bipartition of $X$ iff it satisfies\footnote{Note that when $X_z^{\succeq,\neq}\neq\emptyset$, the condition of $(X_z^{\succeq,\neq}\cup\{z\})\succeq X_z^{\prec}$ in statement (2) of the theorem plays the same role as $X_z^{\succeq,\neq}\succeq X_z^{\prec}$. But when $X_z^{\succeq,\neq}=\emptyset$, the condition of $(X_z^{\succeq,\neq}\cup\{z\})\succeq X_z^{\prec}$ cannot be written as $X_z^{\succeq,\neq}\succeq X_z^{\prec}$.} $(X_z^{\succeq,\neq}\cup\{z\})\succeq X_z^{\prec}$ or $\gamma((X_z^{\succeq,\neq}\cup\{z\}), X_z^{\prec})>0$.
\item [(3)] Provided that $X_z^{\prec}\neq\emptyset$ and $X_z^{\succ}\neq\emptyset$, then the ordered triple $\left \langle X_z^{\succeq,\neq}, \{z\}, X_z^{\prec}\right \rangle$ is a dominating-set-relaxed tripartition of $X$ iff $\beta(X_z^{\succeq,\neq},X_z^{\prec})>0$.
\end{itemize}

\textbf{Proof} The theorem consists of three statements. We will prove them one by one.
\begin{itemize}
\item [(1)] First, because we assume that the binary preference relation is `complete' and $\lvert X \rvert>1$, thus if $X_z^{\prec}=\emptyset$ and $X_z^{\succ}\neq\emptyset$ then we know that $X_z^{\succ}\cup X_z^{\sim,\neq}$ and $\{z\}$ constitute a partition of $X$. Hence the restriction item (1) of Def.1 is satisfied. Second, if $X_z^{\succ}\neq\emptyset$ then we have $\beta(X_z^{\succ}\cup X_z^{\sim,\neq},\{z\})>0$ holds and thus the restriction item (2) of Def.1 is satisfied. Finally, given $X_z^{\succ}\neq\emptyset$ we always have $X_z^{\succ}\cup X_z^{\sim,\neq}\succeq\{z\}$ hence the restriction item (3) of Def.1 is satisfied. Therefore, the sufficiency of the first statement of the theorem is verified.

The necessity of the first statement of the theorem is self-evident according to Def.1.

\item [(2)] The proof is quite similar to that of statement (1). First, because we assume that the binary preference relation is `complete' and $\lvert X \rvert>1$, thus if $X_z^{\prec}\neq\emptyset$ then we know that $X_z^{\succeq,\neq}\cup\{z\}$ and $X_z^{\prec}$ constitute a partition of $X$. Hence the restriction item (1) of Def.1 is satisfied. Second, because $\{z\}\succ X_z^{\prec}$ we thus have $\beta(\{z\}, X_z^{\prec})>0$ hence $\beta((X_z^{\succeq,\neq}\cup\{z\}), X_z^{\prec})>0$, which indicates that the restriction item (2) of Def.1 or item (2) of Def.2 is satisfied. Finally, if $(X_z^{\succeq,\neq}\cup\{z\})\succeq X_z^{\prec}$ or $\beta((X_z^{\succeq,\neq}\cup\{z\}), X_z^{\prec})>0$ then the restriction item (3) of Def.1 or item (3) of Def.2 is satisfied. Therefore, the sufficiency of the second statement of the theorem is verified.

The necessity of the second statement of the theorem is self-evident according to Def.1 and Def.2.

\item [(3)] First, because we assume that the binary preference relation is `complete' and $\lvert X \rvert>1$, thus we know that $X_z^{\succeq,\neq}$, $\{z\}$, and $X_z^{\prec}$ constitute a partition of $X$. Hence the restriction item (1) of Def.3 is satisfied. Second, given $X_z^{\prec}\neq\emptyset$ and $X_z^{\succ}\neq\emptyset$, we always have $X_z^{\succeq,\neq}\succeq\{z\}$ and $\{z\}\succ X_z^{\prec}$, thus we know that the restriction item (2) of Def.3 is satisfied. Finally, given $X_z^{\succ}\neq\emptyset$ we have $\beta(X_z^{\succ},\{z\})>0$ hence  $\beta(X_z^{\succeq,\neq},\{z\})>0$ which indicates that the constraint regarding `$\beta(A,B)$' in item (3) of Def.3 is satisfied, while the constraint for $\left \langle X_z^{\succeq,\neq}, \{z\}, X_z^{\prec}\right \rangle$ regarding `$\beta(A,C)$' in item (3) of Def.3 is none other but the given expression $\beta(X_z^{\succeq,\neq},X_z^{\prec})>0$. Hence the restriction item (3) of Def.3 is satisfied. Therefore, the sufficiency of the third statement of the theorem is verified.

On the other hand, if $\left \langle X_z^{\succeq,\neq}, \{z\}, X_z^{\prec}\right \rangle$ is an ordered tripartition then, according to Def.3, we have $\beta(X_z^{\succeq,\neq},X_z^{\prec})>0$. The necessity of the third statement of the theorem is verified. Q.E.D.
\end{itemize}

Notice that Theorem 1 presents some conditions under which the alternative set could be orderly partitioned with respect to a certain alternative. In other words, the theorem specifies some alternative-specific partitions as defined below.

\textbf{Definition 4} Suppose that a complete binary preference relation $R(X)$ on alternative set $X$ where $1<\lvert X \rvert <+\infty$ is given.
\begin{itemize}
\setlength{\itemsep}{1pt}
\setlength{\parsep}{1pt}
\setlength{\parskip}{1pt}
\item [(1)] For $z\in X$, we say that $X$ can be partitioned into dominating-set-relaxed bipartition $\left \langle X_z^{\succ}\cup X_z^{\sim,\neq}, \{z\}\right \rangle_z$ with respect to $z$ if 
\begin{itemize}
\item [(a)] $X_z^{\prec}=\emptyset$; and
\item [(b)] $X_z^{\succ}\neq\emptyset$.
\end{itemize}
\item [(2)] For $z\in X$, we say that $X$ can be partitioned into dominating-set-relaxed bipartition $\left \langle X_z^{\succeq,\neq}\cup\{z\},X_z^{\prec} \right \rangle_z$ with respect to $z$ if 
\begin{itemize}
\item [(a)]  $X_z^{\prec}\neq\emptyset$; and
\item [(b)]  $(X_z^{\succeq,\neq}\cup\{z\})\succeq X_z^{\prec}$, or, $\gamma((X_z^{\succeq,\neq}\cup\{z\}), X_z^{\prec})>0$.
\end{itemize}
\item [(3)] For $z\in X$, we say that $X$ can be partitioned into dominating-set-relaxed tripartition $\left \langle X_z^{\succeq,\neq}, \{z\}, X_z^{\prec}\right \rangle_z$ with respect to $z$ if
\begin{itemize}
\item [(a)]  $X_z^{\prec}\neq\emptyset$; and
\item [(b)]  $X_z^{\succ}\neq\emptyset$, $\beta(X_z^{\succeq,\neq},X_z^{\prec})>0$.
\end{itemize}
\end{itemize}
\vspace{-0.2cm}
When any of the above instances occurs, we say that $X$ is \textit{divisible with respect to $z$}, and we call the corresponding partitions the \textit{alternative-specific dominating-set-relaxed partitions (for short, alternative-specific DSRPs)}. Moreover, if there exists an alternative with respect to which $X$ can be orderly partitioned, then $X$ is said to be \textit{specifically divisible}. We remark that the subscripts `z' is the alternative-specific index with respect to which the alternative set are partitioned. Specifically, if the alternative set $X$ is specifically divisible with respect to $z$, then we say $X$ is `\textbf{$z$-divisible}', and the corresponding alternative-specific DSRP with respect to $z$ is called `\textbf{$z$-partition}'.
\vspace{0.3cm}

To illustrate, we consider an example taken from Gaertner (2009, Page 110).

\textbf{Example 2} Three voters provide their individual preferences over $X=\{x,y,z,u\}$ as follows (`$\succ_k$' represents voter $k$'s preference)
$$\begin{array}{ll}
 \text{voter 1}: & x \succ_1 y \succ_1 z \succ_1 u,\\
 \text{voter 2}: & y \succ_2 z \succ_2 u \succ_2 x,\\
 \text{voter 3}: & z \succ_3 u \succ_3 x \succ_3 y.
\end{array}
$$
According to the majority rule, the pairwise majority preferences can be obtained as shown by Figure 2.

\begin{figure}[H]
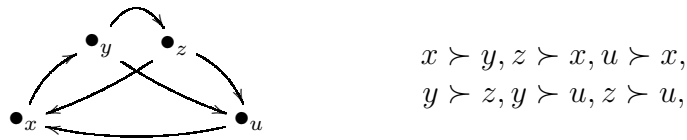

\centering
\subfigure{}
{
\begin{minipage}{5cm}
\[ \xygraph{  
        !{<0cm,0cm>;<1cm,0cm>:<0cm,1cm>::}  
        !{(0,0) }*+{\bullet_{x}}="x"  
        !{(1,1) }*+{\bullet_{y}}="y"  
        !{(2,1) }*+{\bullet_{z}}="z"  
        !{(3,0)}*+{\bullet_{u}}="u"  
            "x":@/^0.2cm/"y"^(0.6){}
            "z":@/^0.1cm/"x"^(0.6){}
            "y":@/_0.1cm/"u" ^(0.8){}   
          "z":@/^0.2cm/"u" ^(0.8){}  
            "u":@/^/"x" ^(0.4){}  
            "y":@/^0.5cm/"z" ^(0.7){}   
        }
\]  
\end{minipage}
}
\subfigure{}
{
\begin{minipage}{5cm}
\[
\begin{array}{ccc}
x\succ y, z\succ x, u\succ x,\\
y\succ z, y\succ u, z\succ u,
\end{array}
\]
\end{minipage}
}
\caption{Pairwise majority relation of Example 2.}
\end{figure}

The pairwise majority relation in this example indicates that each alternative belongs to at least one cycle. However, the alternative set $\{x,y,z,u\}$ is specifically divisible. Even though it cannot be orderly partitioned with respect to $x$ or $y$, it can be orderly partitioned with respect to $z$ and $u$ (the alternative-specific partitions are $\left \langle \{y,z\}, \{x,u\}\right \rangle$$_z$ and $\left \langle \{y,z\}, \{u\}, \{x\}\right \rangle$$_u$, respectively). Hence the alternative set $\{x,y,z,u\}$ in Example 2 is a specifically divisible alternative set. The analysis process is provided in Table 1, where the pairwise majority preferences are included in the left upper corner.

\begin{table}[h]
\renewcommand\arraystretch{1.3}
\centering\footnotesize
\caption{Analysis process of Example 2}
\begin{tabular}{ccccccc}
    \toprule
    \multicolumn{1}{c}{$x\succ y, z\succ x, u\succ x,$}  & \multicolumn{4}{c}{alternative ($a$)} \\
    \cmidrule{2-3}  \cmidrule{4-5}  \cmidrule{6-7}
    {$y\succ z, y\succ u, z\succ u$}  & $x$ & $y$ & $z$ & $u$ \\
    \midrule
$X_a^{\succeq,\neq}$ &

$\{z,u\}
$ &
$\{x\}
$& $\{y\}$ & $\{y,z\}$ \\

$\{a\}$ &

$\{x\}
$ &
$\{y\}
$& $\{z\}$ & $\{u\}$ \\

$X_a^{\prec}$ &

$\{y\}
$ &
$\{z,u\}
$& $\{x,u\}$ & $\{x\}$ \\

$X_a^{\prec}\neq\emptyset?$ &

{Yes} & {Yes} & {Yes} & {Yes} \\

$(X_a^{\succeq,\neq}\cup\{a\})\succeq X_a^{\prec}$? or $\gamma(X_a^{\succeq,\neq}\cup\{a\}, X_a^{\prec})>0$? &

{No} & {No} & {Yes} & {No} \\

${\begin{array}{cc} {\mbox{Can~} X \mbox{~be partitioned into a}}\\
{\mbox{bipartition with respect to~} a?}
\end{array}}$ &

{No} & {No} & {Yes} & {No} \\

$X_a^{\succ}\neq\emptyset$? &

{Yes} & {Yes} & {Yes} & {Yes} \\

$\beta(X_a^{\succeq,\neq},X_a^{\prec})>0$ &

{No} & {No} & {No} & {Yes} \\

${\begin{array}{cc} {\mbox{Can~} X \mbox{~be partitioned into a}}\\
{\mbox{tripartition with respect to~} a?}
\end{array}}$ & 

{No} & {No} & {No} & {Yes} \\

{Alternative-specific partition} &

{--} & {--} & {$\left \langle \{y,z\}, \{x,u\}\right \rangle$$_z$} & {$\left \langle \{y,z\}, \{u\}, \{x\}\right \rangle$$_u$} \\

    \bottomrule
\end{tabular}
\end{table}

In Example 2, every alternative belongs to at least one majority cycle. However, according to Def.3, the alternative set in Example 2 can be orderly partitioned. Therefore, the dominating-set-relaxed partition concept permits some desirable flexibility for dealing with majority cycles.

From the last two parts of Theorem 1 together with Proposition 3, one can see that, when we have $X_z^{\succ}\neq\emptyset$ and $X_z^{\prec}\neq\emptyset$, the alternative set might be partitioned (with respect to alternative $z$) into both a bipartition and a tripartition. This can occur when $X_z^{\succeq,\neq}\succeq X_z^{\prec}$ and $\beta(X_z^{\succeq,\neq},X_z^{\prec})>0$ are additionally satisfied both. An illustration can be the partitions with respect to alternative $b$ in Example 1 as already mentioned in previous subsection. That is, the alternative set in Example 1 can be partitioned with respect to $b$ into a bipartition as $\left \langle \{a,b,c\},\{d\}\right \rangle_b$ and a tripartition as $\left \langle \{a,c\},\{b\},\{d\}\right \rangle_b$. In this situation, the term $b$-partition may refer to any of them.

\subsubsection{Complexity of checking divisibility/outputting specific partitions}
 
From Def.4, we know that whether the alternative set is specifically divisible depends on whether there exists some alternative with respect to which the alternative set can be orderly partitioned. An algorithm (called SeekingPartitions), as shown by Table 2, can thus be designed for checking an alternative set's specific divisibility and outputting the alternative-specific dominating-set-relaxed partitions (if exist). As shown by Theorem 2, it is a polynomial-time algorithm.

\begin{table}[h]
\renewcommand\arraystretch{1.3}
\centering\footnotesize
\caption{Seeking ordered partitions}
\begin{tabular}{llllll}
    \toprule
    \multicolumn{1}{r}{\textbf{Algorithm 1}} & \multicolumn{2}{l} {SeekingPartitions} \\
    \midrule
{ 1: \textbf{Input:}}&{Complete binary preference relation over}&{}\\
                     &{alternative set $X=\{x_1,x_2,\ldots,x_m\}$;}\\
{ 2:}&{divisible:=false;}\\
{ 3:}&{for i:=1 to m}\\
{ }&{\{}\\
{ 4:}&{$X_{x_i}^{\succ}:=\{y\in X\mid y\succ x_i\}$,$X_{x_i}^{\prec}:=\{y\in X\mid y\prec x_i\}$,}\\
{}&{$X_{x_i}^{\sim,\neq}:=\{y\in X\mid y\sim x_i,y\neq x_i\}$,$X_{x_i}^{\succeq,\neq}:=X_{x_i}^{\succ}\cup X_{x_i}^{\sim,\neq}$;}\\
{ 5:}&{if $X_{x_i}^{\prec}=\emptyset$ and $X_{x_i}^{\succ}\neq\emptyset$,} \\
{ }&{then $Par_{x_i}:=\left \langle X_{x_i}^{\succ}\cup X_{x_i}^{\sim,\neq}, \{x_i\}\right \rangle$$_{x_i}$, divisible:=true;}\\
{ 6:}&{else if $X_{x_i}^{\prec}\neq\emptyset$ and \big($(X_{x_i}^{\succeq,\neq}\cup\{x_i\})\succeq X_{x_i}^{\prec}$? or $\gamma(X_{x_i}^{\succeq,\neq}\cup\{x_i\}, X_{x_i}^{\prec})>0$?\big),}\\
{ }&{$~~~~~$ then $ Par_{x_i}:=\left \langle X_{x_i}^{\succeq,\neq}\cup\{x_i\},X_{x_i}^{\prec} \right \rangle$$_{x_i}$, divisible:=true;}\\
{ 7:}&{$~~~~~$ else if $X_{x_i}^{\prec}\neq\emptyset$, $X_{x_i}^{\succ}\neq\emptyset$, and $\beta(X_{x_i}^{\succeq,\neq},X_{x_i}^{\prec})>0$,}\\
{ }&{$~~~~~~~~~~~$ then $ Par_{x_i}:=\left \langle X_{x_i}^{\succeq,\neq}, \{x_i\},X_{x_i}^{\prec} \right \rangle$$_{x_i}$, divisible:=true;}\\
{ }&{\}}\\
{ 8: \textbf{Output:}}&{dominating-set-relaxed partitions w.r.t their each alternative (if divisible=true);} \\
                      &{`None' (otherwise).}
\\
\bottomrule
\end{tabular}
\end{table}

\textbf{Theorem 2} SeekingPartitions (Algorithm 1) runs in polynomial time $O(m^3)$.

\textbf{Proof} Note that the algorithm is iterative, then we start with the operation number within each iteration. Step 4 requires $m-1$ comparisons and $m$ operations for the set union, where $m$ is the number of alternatives. In Step 5, the execution of `$Par_{x_i}:=\left \langle X_z^{\succ}, \{z\}\cup X_z^{\sim,\neq}\right \rangle$$_{x_i}$' requires $m$ operations. Thus it is adequate to execute Step 5 within $m+3$ operations. According to the well-known principle ``when the sum of two positive numbers are a constant, their product is maximal iff they are identical'', we know that `$(X_{x_i}^{\succeq,\neq}\cup\{x_i\})\succeq X_{x_i}^{\prec}$?' and `$\gamma(X_{x_i}^{\succeq,\neq}\cup\{x_i\}, X_{x_i}^{\prec})>0$' in Step6 each requires at most $\frac{1}{4}m^2$ operations. So Step 6 requires no more than $\frac{1}{2}m^2+m+2$ operations. Similarly, Step 7 takes no more than $\frac{1}{4}(m-1)^2+m+3$ operations. Steps 4-7 will repeat $m$ times. By considering the number of iterations, the maximum on the number of operations (after the execution of $m$ iterations) will be $m[(2m-1)+(m+3)+(\frac{1}{2}m^2+m+2)+(\frac{1}{4}(m-1)^2+m+3)]=m(\frac{3}{4}m^2+\frac{9}{2}m+\frac{29}{4})$ in the worst case. We can conclude that Algorithm 1 runs in polynomial time of $O(m^3)$. Q.E.D.

We notice that, Algorithm 1 is designed for seeking alternative-specific partitions in the sense of Def.4 rather than those partitions in the sense of Def.1-Def.3. In addition, when the alternative set can be partitioned into both a bipartition and a tripartition with respect to a certain alternative, the algorithm will output only one of the two partitions, namely the bipartition, as the alternative-specific partition corresponding to the certain alternative. However, such outputs are adequate for our purpose as will be shown in the next section.

\section{Linking alternative-specific DSRP with ranking/selecting}

\subsection{Why and how to link with DSRPs}

Suppose that $R(X)$ is a complete binary preference relation on finite alternative set $X$. At least superficially, it is natural to make connections between the dominating-set-relaxed partitions and choice sets since, for example, a dominating-set-relaxed tripartition $\left \langle A, B, C\right \rangle$ implies $A\succeq B$ and $B\succ C$. Indeed, if the alternative set has a unique tripartition $\left \langle A, B, C\right \rangle$ then we may say that the alternatives in $A$ are `better' than those in $C$. This is a way that is well in line with the Fishburn's Condorcet Rule  `\textit{$x$ is "better than" $y$ iff $x$ beats or ties something that beats $y$}' (Fishburn, 1977). 

However, as demonstrated by Example 1, an alternative set may correspond to a number of dominating-set-relaxed partitions, and an alternative (say, alternative $d$ in Example 1) may occupy a top block in one partition (say, $\left \langle \{a,d\},\{c\},\{b\}\right \rangle$ in Example 1), whereas it may occupy a bottom block in another partition (say, $\left \langle \{a,c\},\{b\},\{d\}\right \rangle$ in Example 1). We cannot claim that an alternative `should' be included in the choice set according to one partition while `should not' according to another. Therefore, to avoid such a confusion we will not employ a `dichotomy' mechanism like `should or should not'. Instead, we will adopt a `soft' scheme in such a way we mean that we assign more or less scores to alternatives in accordance with the ordered blocks they occupy in the dominating-set-relaxed partitions. Whether an alternative is eventually chosen depends on its overall number of scores. The most promising alternatives will be identified as those ones with the highest scores.

In particular, we will employ \textbf{an iterative scheme} in such a way that an alternative may get scores in two ways, namely, the scores obtained as `home points' and the scores as `away points': in each iteration with respect to a certain alternative, if the certain alternative beats some others, then it may get some `home' points, and those alternatives that beat or tie with the certain alternative may get some `away' points. When to assign scores to an alternative, say $x$, based on its comparison with another alternative, say $y$, we will use the following \textbf{scoring formula for a single comparison}
$$\mu(x,y)=\begin{cases}
1, & \mbox{if }x\succ y\\
0.5, & \mbox{if }x\sim y \mbox{ and }x\neq y\\
0, & \mbox{otherwise}
\end{cases},\eqno(5)$$
where the values are attached based upon the following considerations. First, if $x\succ y$ then we can say that $x$ wins one time and thus gets one point from the comparison with $y$, meanwhile $y$ gets no point since it is defeated in the pairwise comparison of this time. Hence it is reasonable to set $\mu(x,y)=1$ if $x\succ y$ and $\mu(x,y)=0$ if $y\succ x$, respectively. Second, we shall not say that one alternative gets some points from the comparison with itself, thus we suppose $\mu(x,x)=0$. Finally, given `$\mu(x,y)=1$ if $x\succ y$, and, $\mu(x,y)=0$ if $y\succ x$', it is possible (not necessary, as will be shown later) to set $\mu(x,y)=0.5$ when $x\sim y$ and $x\neq y$ since we have $x\sim y$ iff $N(x\succ_k y)=N(y\succ_k x)$ under the majority rule.  By the way, we note that our scoring formula for one-time comparison (i.e., formula (5)) is quite similar to a typical definition form of an expression used for computing Copeland scores (Copeland, 1951; Saari and Merlin, 1996; Merlin and Saari, 1997).

\textbf{It should be emphasized that}, in our study whether assign scores to an alternative depends not only on whether it beats some other alternatives but also on some ``particular requirements''. The requirements described in Def.4 are exactly those ones. In fact, it is those conditions that lead toward desired properties of our method (will be detailed in Section 5). We outline our \textbf{scoring strategy} with those conditions as follows.
\begin{itemize}
\item [(1)] We adopt an iterative scheme for assigning possible scores to alternatives. The procedure is done iteratively for each alternative in such a manner that the iteration can be specified by alternatives.
\item [(2)] In the iteration with respect to a certain alternative, say $z$, let $score_z(x)$ denote alternative $x$'s score obtained in the iteration specified by alternative $z$.
\begin{itemize}
\item [a)] if the certain alternative beats none of the other alternatives ($X_z^{\prec}=\emptyset$), then it does not get any `home' score: $score_z(z)=0$, if $X_z^{\prec}=\emptyset$. 

However, those alternatives that beat or tie the certain alternative will get some `away' scores on the condition that there exists at least one alternative that beats the certain alternative: $score_z(x)=\mu(x,z)$, if $X_z^{\succ}\neq\emptyset$ and $x\in X_z^{\succ}\cup X_z^{\sim,\neq}$.
\item [b)] if the certain alternative beats some of the other alternatives ($X_z^{\prec}\neq\emptyset$), then, it will get some `home' scores and those alternatives (if any) that beat or tie the certain alternative will get some `away' scores $score_z(x)=\mu(x,z)+\sum\limits_{y\in X_z^{\prec}}\mu(x,y)$, for $x\in \{z\}\cup X_z^{\sim,\neq}\cup X_z^{\succ}$, on the condition that one of the following holds
\begin{itemize}
\item each one of those alternatives that beat or tie the certain alternative is not beaten by any of those alternatives that are beaten by the certain alternative (i.e., $(X_z^{\succeq,\neq}\cup\{z\})\succeq X_z^{\prec}$). 
\item each one of those alternatives that beat or tie the certain alternative beats at least one of those alternatives that are beaten by the certain alternative (i.e., $\gamma(X_z^{\succeq,\neq}\cup\{z\}, X_z^{\prec})>0$).
\item the certain alternative is beaten by some of the others (i.e., $X_z^{\succ}\neq\emptyset$), and, there exists at least one alternative distinct from $z$, denoted by $z^*$, such that $z^*$ not only beats or ties with the certain alternative but also beats all those alternatives that are beaten by the certain alternative (i.e., $\beta(X_z^{\succeq,\neq}, X_z^{\prec})>0$).
\end{itemize}
In addition, those alternatives that are beaten by the certain alternative will get no `away' score, i.e., $score_z(x)=0$ if $x\in X_z^{\prec}$.
\end{itemize}
\item [(3)] The total score of alternative $x$, denoted by $\psi(x)$, is obtained by racking up the points that the alternative receives from each iteration, namely, $\psi(x)=\sum_{z\in X}score_z(x)$. 
\end{itemize}

By making a comparison between the above `scoring strategy' and Def.4, the reader may have observed their connections. In fact, the scoring strategy states how to use the alternative-specific dominating-set-relaxed partitions to assign scores. We'd like to emphasize two elements of the strategy. On one hand, the following four kinds of alternative-specific dominating-set-relaxed partitions with respect to an alternative, say $z$, will probably be used.
\begin{itemize}
\item [k1:] The bipartition $\langle X_{z}^{\succeq,\neq},\{z\} \rangle_{z}$ if $X_{z}^{\prec}=\emptyset$ and $X_{z}^{\succ}\neq\emptyset$;
\item [k2:] The bipartition $\langle X_{z}^{\succeq,\neq}\cup\{z\}, X_{z}^{\prec}\rangle_{z}$ if $X_{z}^{\prec}\neq\emptyset$ and $(X_z^{\succeq,\neq}\cup\{z\})\succeq X_z^{\prec}$;
\item [k3:] The bipartition $\langle X_{z}^{\succeq,\neq}\cup\{z\}, X_{z}^{\prec}\rangle_{z}$ if $X_{z}^{\prec}\neq\emptyset$ and $\gamma(X_z^{\succeq,\neq}\cup\{z\}, X_z^{\prec})>0$; and
\item [k4:] The tripartition $\langle X_{z}^{\succeq,\neq},\{z\}, X_{z}^{\prec}\rangle_{z}$ if $X_{z}^{\prec}\neq\emptyset$, $X_{z}^{\succ}\neq\emptyset$ and $\beta(X_z^{\succeq,\neq}, X_z^{\prec})>0$.
\end{itemize}
On the other hand, when to assign scores, the alternatives contained in the bottom blocks will get 0 scores, while those alternatives that are included in the middle and top blocks will get some positive `home' scores or positive `away' scores. 

Although the alternative-specific partitions (as defined by Def.4) are exclusively used when assigning scores, however, as will be shown in Section 5, such a scoring strategy leads to a ranking/choosing procedure with a number of appealing properties. Therefore, the alternative-specific partitions are adequate for our study in this paper.

\subsection{Scoring function, algorithm and complexity}

\subsubsection{Formula for computing alternative scores}

If we assume $\sum\limits_{y\in X_z^{\prec}}\mu(x,y)$ to be 0 when $X_z^{\prec}=\emptyset$, then the scoring strategy outlined in subsection 4.1 can be formulated as a \textbf{scoring rule}
$$\psi(x)=\sum_{z\in X}score_z(x), \forall x\in X,\eqno(6)$$
where
\begin{footnotesize}
$$score_z(x)=
\begin {cases} 
\mu(x,z)+\sum\limits_{y\in X_z^{\prec}}\mu(x,y),&\text{if one of the conditions~} {\scriptsize\begin{Bmatrix}\big(X_z^{\prec}=\emptyset\text{~and~}X_z^{\succ}\neq\emptyset\text{~and~}x\in X_z^{\succeq,\neq}\big), \\ \big(X_z^{\prec}\neq\emptyset\text{~and~}Z_z^{\succeq}\succeq X_z^{\prec}\text{~and~}x\in X_z^{\succeq}\big), \\ \big(X_z^{\prec}\neq\emptyset\text{~and~}\gamma(Z_z^{\succeq}, X_z^{\prec})>0\text{~and~}x\in X_z^{\succeq}\big), \\ \big(X_z^{\prec}\neq\emptyset\text{~and~}X_z^{\succ}\neq\emptyset\text{~and~}\beta(X_z^{\succeq,\neq},X_z^{\prec})>0\text{~and~}x\in X_z^{\succeq}\big) \end{Bmatrix}}\text{~holds},\\ 
0,&\text{else.} 
\end {cases}\eqno(7)$$
\end{footnotesize}

\vspace{0.2cm}

Alternative $x$'s final score, denoted $\psi(x)$, is the sum of those scores, denoted $score_z(x), z\in X$, which are derived from checking each alternative's possible specific partitions. Only when the alternative set $X$ is divisible with respect to $z$ and $x$ is not contained in the bottom block of the alternative-specific DSRP with respect to $z$ can $score_z(x)$ be a nonzero value. Therefore, \textbf{the value of $score_z(x)$ can be regarded as the score that alternative $x$ gains from alternative $z$}, which can be rewritten as
\begin{footnotesize}
$$score_z(x)=
\begin {cases} 
\mu(x,z),&\text{if~} X_z^{\prec}=\emptyset, \langle X_{z}^{\succeq,\neq},\{z\} \rangle_{z}\text{~is the~} z\text{-partition},\text{~and~}x\in X_z^{\succeq,\neq},\\ 
\mu(x,z)+\sum\limits_{w\in X_z^{\prec}}\mu(x,w),&\text{else if~} X_{z}^{\prec}\neq\emptyset, \text{~either of~} {\scriptsize\begin{Bmatrix}\langle X_{z}^{\succeq,\neq}\cup\{z\}, X_{z}^{\prec}\rangle_{z}, \\ \langle X_{z}^{\succeq,\neq},\{z\}, X_{z}^{\prec}\rangle_{z} \end{Bmatrix}}\text{~is the~} z\text{-partition~},\text{~and~}x\in X_z^{\succeq},\\ 
0,&\text{~~~~~~else.} 
\end {cases}$$
\end{footnotesize}

Moreover, if we introduce a $\delta$-function in the following fashion
$$\delta(``restrictions'')=\begin{cases}
1, & \mbox{if the ``\textit{restrictions}'' are satisfied,}\\
0, & \mbox{otherwise,}
\end{cases}\eqno(8)$$
then the above scoring rule can be rewritten as
\begin{footnotesize}
$$\begin{array}{lll}
\psi(x)=&\sum\limits_{z\in X}\Big[\delta\big(X_z^{\prec}=\emptyset,X_z^{\succ}\neq\emptyset\big)\delta(x\in X_z^{\succeq,\neq})\mu(x,z)+\\
&\delta(X_z^{\prec}\neq\emptyset)\delta\Big(Z_z^{\succeq}\succeq X_z^{\prec} \lor \gamma(Z_z^{\succeq}, X_z^{\prec})>0 \lor \big(X_z^{\succ}\neq\emptyset,\beta(X_z^{\succeq,\neq},X_z^{\prec})>0\big)\Big)\delta(x\in X_z^{\succeq})\big(\mu(x,z)+\sum\limits_{y\in X_z^{\prec}}\mu(x,y)\big)\Big].
\end{array}\eqno(9)$$
\end{footnotesize}

\vspace{0.2cm}

At this point we shall attach a special emphasis to one aspect of our scoring strategy. As mentioned earlier (subsection 3.4.2), an alternative may correspond to both a specific bipartition and a specific tripartition. When to assign scores, one and only one of them is used so as to avoid duplicate calculations. In fact, as can be seen from expressions (7) and (9), either of them (if exist) can be employed for the score calculation. In this situation, Algorithm 1 will output the specific bipartition alone. The rationality lies in the fact that, when the alternative set can be partitioned, with respect to alternative $z$, into both a specific bipartition
$$\left \langle X_z^{\succeq,\neq}\cup\{z\},X_z^{\prec} \right \rangle_z$$
and a specific tripartition
$$\left \langle X_z^{\succeq,\neq}, \{z\}, X_z^{\prec}\right \rangle_z,$$
our scoring strategy will ensure that the specific bipartition and the specific tripartition make equal contributions to the score of each alternative. That is, when a specific partition is used to assign scores, the alternatives in the bottom block will be assigned zero values, while the others will be assigned non-zero values in a similar way no matter what kind of partition it is. This can be easily checked by testing Expression (9) on the above two specific partitions.

\subsubsection{Algorithm for computing alternative scores and complexity}

Given a collection of pairwise comparison outcomes among a finite set of alternatives, one may like to implement the computation of the alternative scores on a computer. We present a sketched version of an algorithm (called ComputingScores) based on expression (9), as shown by Table 3. 

\begin{table}[h]
\renewcommand\arraystretch{1.3}
\centering\footnotesize
\caption{Computing alternative scores}
\begin{tabular}{llllll}
    \toprule
    \multicolumn{1}{r}{\textbf{Algorithm 2}} & \multicolumn{2}{l} {ComputingScores} \\
    \midrule
{ 1: \textbf{Input:}}&{Complete binary preference relation over}&{}\\
                     &{alternative set $X=\{x_1,x_2,\ldots,x_m\}$;}\\
{ 2:}&{for i:=1 to m}\\
{   }&\{$\psi(x_i):=0$;\}\\
{ 3:}&{for i:=1 to m}\\
{ }&{\{}\\
{ 4:}&{$X_{x_i}^{\succ}:=\{y\in X\mid y\succ x_i\}$,$X_{x_i}^{\prec}:=\{y\in X\mid y\prec x_i\}$,}\\
{}&{$X_{x_i}^{\sim,\neq}:=\{y\in X\mid y\sim x_i,y\neq x_i\}$,$X_{x_i}^{\succeq,\neq}:=X_{x_i}^{\succ}\cup X_{x_i}^{\sim,\neq}$;}\\
{ 5:}&{if $X_{x_i}^{\prec}=\emptyset$ and $X_{x_i}^{\succ}\neq\emptyset$,} \\
{ }&{then for j:=1 to m}\\
{ }& \hspace{0.4cm} \{if $x_j\in X_{x_i}^{\succeq,\neq}$, then $\psi(x_j):=\psi(x_j)+\mu(x_j,x_i)$;\}\\
{ 6:}&{else if $X_{x_i}^{\prec}\neq\emptyset$ and \big($(X_{x_i}^{\succeq,\neq}\cup\{x_i\})\succeq X_{x_i}^{\prec}$? or $\gamma(X_{x_i}^{\succeq,\neq}\cup\{x_i\}, X_{x_i}^{\prec})>0$?\big),}\\
{ }&\hspace{0.4cm}  {then for j:=1 to m}\\
{ }& \hspace{0.8cm} \{if $x_j\in (X_{x_i}^{\succeq,\neq}\cup\{x_i\})$, then $\psi(x_j):=\psi(x_j)+\mu(x_j,x_i)+\sum\limits_{y\in X_{x_i}^{\prec}}\mu(x_j,y)$;\}\\
{ 7:}&{\hspace{0.4cm}  else if $X_{x_i}^{\prec}\neq\emptyset$, $X_{x_i}^{\succ}\neq\emptyset$, and $\beta(X_{x_i}^{\succeq,\neq},X_{x_i}^{\prec})>0$,}\\
{ }&\hspace{1.0cm}  {then for j:=1 to m}\\
{ }& \hspace{1.4cm} \{if $x_j\in (X_{x_i}^{\succeq,\neq}\cup\{x_i\})$, then $\psi(x_j):=\psi(x_j)+\mu(x_j,x_i)+\sum\limits_{y\in X_{x_i}^{\prec}}\mu(x_j,y)$;\}\\
{ }&{\}}\\
{ 8: \textbf{Output:}}&{$\psi(x_i), x_i\in X$;} \\
\\
\bottomrule
\end{tabular}
\end{table}

By making a comparison between Table 2 and Table 3, one can see that Algorithm 2 is obtained from Algorithm 1 by substituting several iterations for several assignment operations. Such substitution does not make the complexity of the algorithm worse. Therefore, we have the following corollary deduced from Theorem 2.

\textbf{Corollary 1} ComputingScores (Algorithm 2) runs in polynomial time $O(m^3)$.

\vspace{0.2cm}

\section{Possible application scenarios and properties}

The expression (9) (or Algorithm 2) represents a procedure for assigning scores to alternatives so as to obtain an ordering over the alternative set and then select a single winner (the alternative receiving the highest score) or a size-$k$ committee (those alternatives receiving the $k$ highest scores). Because the procedure is based solely on the pairwise comparisons between alternatives as input without the prerequisite for transitive relations or the prohibition of ties, the proposed procedure may be suitable for a wide variety of applications with different kinds of input data. For the sake of convenience, in what follows the proposed scoring function (i.e., expressions (6)-(9)) will be referred to as ``\textbf{the DSR-scoring method}'' as a result of the fact that it asssigns scores by means of the dominating-set-relaxed partitions (DSRPs). Accordingly, the score of an alternative that is computed by the DSR-scoring method will be called ``\textbf{DSR-score}'' or, more briefly, score. In addition, sometimes we will add `a' subscript to $\psi(x)$ to identify the inupt data based on which the alternatives' DSR-scores are calculated. For instance, $\psi_{M(X)}(x)$, $\psi_{\varrho(X)}(x)$ and $\psi_{T(X)}(x)$ stand for that the DSR-scores are calculated based on a pairwise majority relation $M(X)$, a profile $\varrho(X)$ and a tournament $T(X)$, respectively. Of course, when the context is trivially clear, or the outcome is context-independent, we generally omit this kind of subscript for conciseness.

\vspace{2mm}

The following lemma will be useful in subsequent proofs.

\textbf{Lemma 1} Suppose that $O(X)$ is an ordering on the alternative set $X$, and let $O(X)$ be the input of the DSR-scoring method. If $[\forall (x,y)\in X\times X: x\sim y]$, then $[\forall x\in X: \psi(x)=0]$; else, then $[\forall (x,y)\in X\times X: x\succ y\rightarrow \psi(x)>\psi(y)]$ and $[\forall (x,y)\in X\times X: x\sim y\rightarrow \psi(x)=\psi(y)]$.

\textbf{Proof} (I) If $[\forall (x,y)\in X\times X: x\sim y]$, then the alternative set $X$ is indivisible with respect to any alternative since, by Def.4 we know that a necessary condition for $X$ to be specifically divisible is that their exists at least one pair $(x,y)\in X\times X$ such that either $x\succ y$ or $y\succ x$, but not both. Furthermore, by expression (9) we know that all alternatives will be assigned zero values when the alternative set is NOT specifically divisible. That is, in this situation where $[\forall (x,y)\in X\times X: x\sim y]$, we have $[\forall x\in X: \psi(x)=0]$.

(II) Else (i.e., $\exists (x,y)\in X\times X: x\succ y$), we will prove the statement from two respects given the input of the DSR-scoring method is an ordering. On one hand, it will be asserted that the alternative set $X$ is specifically divisible with respect to each alternative. On the other hand, we will use expression (9) for calculating scores. 

Suppose that $(x,y)\in X\times X$ and $x\succ y$. Then: 
\begin{itemize}
\item [(i)] we have $X_x^{\prec}\neq\emptyset$ since $X_x^{\prec}$ contains at least $y$. By virtue of the transitivity property of ordering, we can assert $(X_x^{\succeq,\neq}\cup\{x\})\succ X_x^{\prec}$. By Def.4, we know that $\langle X_x^{\succeq,\neq}\cup\{x\}, X_x^{\prec} \rangle_{x}$ is an $x$-partition\footnote{Note that the introduced `specific bipartition' arises as an integral part in our `designing' a SWF/SCC. In addition, here $\langle X_x^{\succeq,\neq}\cup\{x\}, X_x^{\prec} \rangle_{x}$ can be viewed as either a first-type bipartition or a second-type bipartition by Def.1, Def.2 and Def.4.}. By our scoring strategy or expression (9), we can further assert that $score_x(x)>score_x(y)=0$ as a result of the fact that $x$ is in the top block of $x$-partition, while $y$ is in the bottom block; 
\item [(ii)] we have $X_y^{\succ}\neq\emptyset$ since $X_y^{\succ}$ contains at least $x$. If $X_y^{\prec}=\emptyset$, by Def.4 we then know that $\langle X_y^{\succeq,\neq}, \{y\} \rangle_{y}$ is a $y$-partition\footnote{Note that the `specific bipartition' defined by Def.4(1) arises as an integral part in our `designing' a SWF/SCC.}. Since $x\in X_y^{\succ}$ hence $x\in X_y^{\succeq,\neq}$ and thus $x$ is in the top block. Meanwhile, since $y$ is in the bottom block, we thus have $score_y(x)>score_y(y)$ in the case where $X_y^{\prec}=\emptyset$. If $X_y^{\prec}\neq\emptyset$, since $x\succ y$, and by transitivity, we have $\{x\}\succ X_y^{\prec}$ and $X_y^{\succeq,\neq}\succ X_y^{\prec}$. Furthermore, by Def.4 we know that $\langle X_y^{\succeq,\neq}, \{y\}, X_y^{\prec} \rangle_{y}$ is a $y$-partition\footnote{Here, the blocks of $X_y^{\succeq,\neq}$, $\{y\}$ and $X_y^{\prec}$ can also constitute of a specific bipartition with respect to $y$, namely, $\langle X_y^{\succeq,\neq}\cup \{y\}, X_y^{\prec} \rangle_{y}$. It is worth noting that, by expression (9), here the tripartition $\langle X_y^{\succeq,\neq}, \{y\}, X_y^{\prec} \rangle_{y}$ and the bipartition $\langle X_y^{\succeq,\neq}\cup \{y\}, X_y^{\prec} \rangle_{y}$ will produce the same values of $score_y(x)$ and $score_y(y)$. Therefore, we can make use of either of them and the final result will keep the same.}, where $x$ is in the top block, while $y$ is in the middle block. Since we already obtained $\{x\}\succ X_y^{\prec}$ and we are given $x\succ y$, by expression (9) we obtain $score_y(x)=1+|X_y^{\prec}|$ and $score_y(y)=|X_y^{\prec}|$. We thus have $score_y(x)>score_y(y)$ in the case where $X_y^{\prec}\neq\emptyset$. In this step, we obtain $score_y(x)>score_y(y)$ no matter whether or not $X_y^{\prec}$ is $\emptyset$.

At this stage, in fact, we have validated a reduced version of the lemma when used with $|X|=2$. Here the term `reduced' means when $|X|=2$ we only need to show that the statement of Lemma 1 is valid for $x\succ y$ and $x\sim y$.
\item [(iii)] Let $z$ be any alternative distinct from $x$ and $y$ in a situation where $|X|>2$. Due to completeness, there is one and only one possible relationship between $y$ and $z$ in the input data, that is, $y\succ z$, $y\sim z$ or $z\succ y$.
\begin{itemize}
\item [(a)] When $y\succ z$, since $x\succ y$, and, by transitivity, we have $x\succ z$. It can be asserted, in the same manner as an earlier step above (i.e., step II.ii), that $X$ is specifically divisible with respect to $z$, and that both $x$ and $y$ are contained in the top block of the resultant $z$-partition.\footnote{It may be a bipartition or a tripartition.} Consequently, by expression (9) we obtain that $score_z(x)=score_z(y)=1+|X_z^{\prec}|$.
\item [(b)] When $y\sim z$, since $x\succ y$, and, by transitivity, we also have $x\succ z$. By the same reasoning as before (i.e., step II.ii), we know that $X$ is specifically divisible with respect to $z$, and that both $x$ and $y$ are contained in the top block of the resultant $z$-partition. It is noteworthy that in the current situation we are given $y\sim z$ rather than $y\succ z$. Therefore, by expression (9) we obtain $score_z(x)=1+|X_z^{\prec}|$ and $score_z(y)=\frac{1}{2}+|X_z^{\prec}|$, so that we have $score_z(x)>score_z(y)$ in the situation where $y\sim z$. By the way, since $y\sim z$, we know that $z$ is included in the top block of $y$-partition, and, by transitivity we have $X_y^{\prec}=X_z^{\prec}$, so that we have $score_y(z)=\mu(z,y)+\sum_{w\in X_y^{\prec}}\mu(z,w)=\mu(y,z)+\sum_{w\in X_z^{\prec}}\mu(y,w)=score_z(y)=\frac{1}{2}+|X_z^{\prec}|$.
\item [(c)] When $z\succ y$, since $x\succ y$, we distinguish three cases:
\begin{itemize}
\item [--] when $z\succ y$ and $z\prec x$, we assert, in a similar way to an earlier step above (i.e., step II.i), that $X$ is specifically divisible with respect to $z$, and that $x$ is contained in the top block of the resultant $z$-partition, while $y$ is in the bottom block. Consequently, by expression (9) we know that $score_z(x)>score_z(y)=0$.
\item [--] when $z\succ y$ and $z\sim x$, as analyzed in an earlier step (II.i in this proof), we know that $X$ is specifically divisible with respect to $z$ since $z\succ y$. Moreover, $x$ is contained in the top block of the resultant $z$-partition, while $y$ is in the bottom block. We thus have $score_z(x)>score_z(y)=0$. By the way, since $z\sim x$, we know that $z$ is included in the top block of $x$-partition, and, by transitivity we have $X_x^{\prec}=X_z^{\prec}$, so that we have $score_x(z)=score_z(x)$.
\item [--] when $z\succ y$ and $z\succ x$, as asserted in II.i of this proof, we know that $X$ is specifically divisible with respect to $z$, and $z$ will be contained in the top block of the resultant $z$-partition, while both $x$ and $y$ are in the bottom block, so that we have $score_z(x)=score_z(y)=0$. 
\end{itemize}
\end{itemize}
From the steps II.i-iii in this proof we infer that $\psi(x)=\sum_{w\in X}score_w(x)>\sum_{w\in X}score_w(y)=\psi(y)$ when $x\succ y$, since $score_x(x)>score_x(y)$ and $\forall w\in X\setminus\{x\}[score_w(x)\geq score_w(y)]$. In addition, since in the steps II.i-iii in this proof we asserted that $X$ is not only $x$-divisible, but also $y$-divisible, as well as $z$-divisible for any $z$ distinct from $x$ and $y$, thus we deduce that \textit{$X$ is specifically divisible with respect to each alternative, provided that the preference relation is an ordering and their exists an alternative pair $(x,y)$ such that $x\succ y$}. 
\item [(iv)] It will be shown that $u\sim v$ leads to $\psi(u)=\psi(v)$ under our hypotheses. First we know $|X|>2$, since we are given $x\succ y$ and, in the current step, one of our hypotheses is $u\sim v$. Let $z$ be an alternative distinct from $u$ and $v$. And then note that we have asserted that $X$ is specifically divisible with respect to each alternative in our context where $O(X)$ is an ordering and we are given $\exists (x,y)\in X\times X(x\succ y)$.
\begin{itemize}
\item [--] As we asserted in Step II.iii.b of this proof, we have $score_u(v)=score_v(u)$ provided $u\sim v$.
\item [--] Since the preference relation is an ordering (complete and transitive), by the hypothesis $u\sim v$, we know that: 
\begin{itemize}
\item [(a)] $X_u^{\prec}=X_v^{\prec}$ hence $score_u(u)=|X_u^{\prec}|=|X_v^{\prec}|=score_v(v)$; 
\item [(b)] $z\succ u$ iff $z\succ v$; $z\sim u$ iff $z\sim v$; $z\prec u$ iff $z\prec v$; and thus $\{u\}\succ X_z^{\prec}$ iff $\{v\}\succ X_z^{\prec}$. Since from steps II.i-iii in this proof we have asserted that the alternative set $X$ is specifically divisible with respect to each alternative, hence $X$ is specifically divisible with respect to $z$. Therefore, if $u\succeq z$ then both $u$ and $v$ are in the top block of $z$-partition, so that we have $score_z(u)=\mu(u,z)+\sum_{w\in X_z^{\prec}}\mu(u,w)=\mu(v,z)+\sum_{w\in X_z^{\prec}}\mu(v,w)=score_z(v)$; else (i.e., $z\succ u$) then both $u$ and $v$ are in the bottom block of $z$-partition and we thus have $score_z(u)=score_z(v)=0$.
\end{itemize}
\end{itemize}
It follows that $\psi(u)=score_u(u)+score_v(u)+\sum_{w\in X\setminus\{u,v\}}score_w(u)=score_u(v)+score_v(v)+\sum_{w\in X\setminus\{u,v\}}score_w(v)=\psi(v)$ under conditions that $O$ is an ordering, $u\sim v$ and $\exists (x,y)\in X\times X(x\succ y)$, so that the lemma holds. Q.E.D.
\end{itemize}

\vspace{2mm}

\textbf{Remark 1} In the proof of Lemma 1, if the point to be assigned to both sides of a tie changes from 0.5 to another value from $[0,1]$, then the statement still holds.

\vspace{2mm}

It is immediately clear that Lemma 1 has a simplified version of itself: 

\textbf{Lemma 1'} If the input of the DSR-scoring method is an ordering $O(X)$ on the alternative set $X$, then for all alternative pair $(x,y)$, we have $x\succ y\rightarrow \psi(x)>\psi(y)$, and $x\sim y\rightarrow \psi(x)=\psi(y)$.

\vspace{2mm}

Moreover, from the proof of Lemma 1 we have the following two corollaries.

\textbf{Corollary 2} Suppose that the preference relation over a finite alternative set is an ordering, and their exists an alternative pair such that one is preferred to the other. Then it follows that the alternative set is specifically divisible with respect to each alternative.

\vspace{2mm}

\textbf{Corollary 3} (\textbf{Scoring formula for linear ordering}) If the input data indicate a linear ordering of the alternatives, then the DSR-score of the alternative in the $i$-last position (the one-last being the least preferred) is\footnote{In other words, the DSR-scoring method assigns a score of $\frac{1}{2}(m-i)(m+3-i)$ to an alternative every time it is $i$th most preferred by a simple majority.} 
$$\psi(x_{i\mbox{-last}})=\frac{1}{2}(i-1)(i+2),\eqno(10)$$
which suggests scores for the alternatives from least preferred to most preferred as 0, 2, 5, 9, $\cdots$, $\frac{1}{2}(i-1)(i+2)$, $\cdots$.\footnote{One may have observed the relationship between the well-known Borda points $\ss_{i\mbox{-last}}$ and the DSR-score $\psi(x_{i\mbox{-last}})$ when the alternatives are ranked into a linear order, that is, $\ss_{i\mbox{-last}}=\frac{\psi(x_{i\mbox{-last}})}{\frac{1}{2}(i+2)}$, where $\ss_{i\mbox{-last}}$ refers to the Borda point sequence 0, 1, 2, 3, $\cdots$, $i-1$, $\cdots$.}

It is noteworthy that, when the input data of the DSR-scoring method constitute a linear order (without ties), the derived alternative DSR-scores are none other than the triangular numbers minus 1.\footnote{Triangular numbers refer to the number sequence 1, 3, 6, 10, 15, $\ldots$, $\mathcal{T}_t$, $\ldots$, where $\mathcal{T}_t=1+2+\ldots+t=\frac{1}{2}t(t+1)$ (Castillo, 2016; Tattersall, 1999). This type of figurate number was initially studied by Pythagoras (a well-known mathematician and philosopher in Ancient Greece) and his followers.}

\subsection{When used as a SWF/SCC}

When the individual preferences are reported as orderings, the DSR-scoring method can be used as a SWF (i.e., a mapping from preference profiles to alternative orderings) or a SCC (i.e., a mapping from preference profiles to nonempty sets of alternatives).  
\begin{itemize}
\item [] In this scenario (given a preference profile $\varrho(X)$), we first deduce the simple majority relation on all the pairs of alternatives from the individuals' orderings, and then we use the DSR-scoring method to obtain the alternatives' scores $\psi_{\varrho(X)}(x)$ based on which we can derive an ordinal ranking of the alternatives (The higher the score, the higher the ranking of an alternative). And then, the choice set will be $DSR(\varrho(X))=\arg\max_{x\in X}\psi_{\varrho(X)}(x)$ (i.e., those alternatives with maximal DSR-scores). 
\end{itemize}

We have the following resluts.

\textbf{Theorem 3} When the DSR-scoring method is used as a SWF, it has the following properties:
\begin{itemize}
\item [(1)] As a SWF, the DSR-scoring method satisfies the weak Pareto principle (WP).
\item [(2)] As a SWF, the DSR-scoring method satisfies the strong Pareto principle (SP).
\item [(3)] As a SWF, the DSR-scoring method satisfies the property of non-dictatorship (ND).
\item [(4)] As a SWF, if the majority relation induced by any profile in its domain is an ordering on $X$, then the DSR-scoring method satisfies the property of independence of irrelevant alternatives (IIA). 
\end{itemize}

\textbf{Proof}\begin{itemize}
\item [(1)] \textbf{(WP)} Under the condition that the individuals' preferences are orderings (i.e., complete and transitive over $X$), we need to prove that $\forall x,y \in X$, if $\forall k(x\succ_k y)$ then $\psi(x)>\psi(y)$. 

We start off by showing that $\forall z(y\succ_M z\rightarrow x\succ_M z)$\footnote{Miller (2013) asserted this expression when it is for a tournament under the hypothesis of $\forall k(x\succ_k y)$. I quote here, ``\textit{if x is unanimously preferred to y, then x is majority preferred to all alternatives to which y is majority preferred}''.} and $\forall z(y\sim_M z\rightarrow x\succeq_M z)$, where the subscript `$M$' indicates `by pairwise majority decision'. Obviously, we trivially have $x\succ_M y$ by condition $\forall k(x\succ_k y)$. In the next step, we point out that: (a) the alternative set $X$ can always be divisible with respect to $y$ so that alternative $x$ will get more scores from $y$ than $y$ gets from itself; and (b) alternative $x$ will get score from alrenative $z$ other than $y$ at least as high as the score that $y$ gets from $z$ no matter whether or not the alternative set $X$ is divisible (with respect to $z$). Then the statement of WP will be validated.

To show $\forall z(y\succ_M z\rightarrow x\succ_M z)$ and $\forall z(y\sim_M z\rightarrow x\succeq_M z)$, we examine what can be deduced from two assumptions we have made. Since we assumed that all the individuals' preferences are orderings (complete and transitive) and that  $\forall k(x\succ_k y)$, by the interpretation of `transitive' in Section 2 we have $\forall k(y\succ_k z\rightarrow x\succ_k z)$ and $\forall k(y\sim_k z\rightarrow x\succ_k z)$. Accordingly, we obtain $\forall z(y\succ_M z\rightarrow x\succ_M z)$ and $\forall z(y\sim_M z\rightarrow x\succeq_M z)$. The latter deduction may not be so obvious, so that we give an interpretation: in a situation where $N(y\succ_k z)=N(z\succ_k y)=\frac{n}{2}$ and $\forall k(z\succ_k y\rightarrow z\succ_k x)$, we will have $y\sim_M z\rightarrow x\sim_M z$; in other situations we will have $y\sim_M z\rightarrow x\succ_M z$. Therefore we obtain $\forall z(y\sim_M z\rightarrow x\succeq_M z)$.

Now we focus on the scores of $x$ and $y$, namely, $\psi(x)=\sum_{z\in X}score_z(x)$ and $\psi(y)=\sum_{z\in X}score_z(y)$. As mentioned in subsection 4.2, $score_z(x)$ is the score that $x$ gains from $z$: it will take a positive value if $X$ is divisible with respect to $z$ and if $x$ is NOT in the bottom block; otherwise, it will be zero. We first show $score_y(x)>score_y(y)$, then we show $score_z(x)\geq score_z(y), \forall z\in X\setminus \{y\}$. 

We will have $score_y(x)>score_y(y)$, since: if $X_y^{\prec}=\emptyset$, then $\langle X_{y}^{\succeq,\neq},\{y\} \rangle_{y}$ is the $y$-partition since $x\succ_M y$ hence $X_y^{\succ}\neq\emptyset$ and thus the statement (1) of Def.4 is satisfied. In this situation, since $y$ is in the bottom block, thus $score_y(y)=0$; meanwhile, since $x$ is in the top block resulting from $x\succ_M y$, by formula (7) we thus have $score_y(x)=1$. Therefore, we have $score_y(x)>score_y(y)$; else (i.e., $X_y^{\prec}\neq\emptyset$), for every $z\in X_y^{\prec}$, we have $y\succ_M z$. As already shown in the case of $y\succ_M z$ we will have $x\succ_M z$ as well. Hence we have $\beta(X_y^{\succeq,\neq},X_y^{\prec})>0$ and thus $\langle X_{y}^{\succeq,\neq},\{y\},X_y^{\prec} \rangle_{y}$ is the $y$-partition\footnote{Note that the `specific tripartition' defined by Def.4(3) arises as an integral part in our `designing' a SWF/SCC.}. In this situation by formula (7) we have $score_y(y)=|X_y^{\prec}|$ and $score_y(x)=1+|X_y^{\prec}|$ and thus $score_y(x)>score_y(y)$. Therefore, no matter whether or not $X_y^{\prec}$ is $\emptyset$, the alternative set $X$ can always be divisible with respect to $y$ and we always have $score_y(x)>score_y(y)$.

We will have $score_z(x)\geq score_z(y), \forall z\in X\setminus \{y\}$, since: 
\begin{itemize}
\item [(I)] given $z\in X\setminus \{y\}$, if $X$ is not divisible with respect to $z$, then, according to formula (7) or (9) we know that $score_z(x)=score_z(y)=0$. In this situation we thus have $score_z(x)\geq score_z(y)$; 
\item [(II)] else (i.e., $X$ is divisible with respect to $z$, where $z\in X\setminus \{y\}$), we distinguish two cases: 
\begin{itemize}
\item [(a)] given $z\neq y$ and $X$ is divisible with respect to $z$, if $z=x$ then $X$ is divisible with respect to $x$ (since $z=x$ and our hypothesis is that $X$ is divisible with respect to $z$). In this situation alternative $y$ must be included in the bottom block of the $x$-partition since $x\succ_M y$. By formula (7) we thus have $score_x(y)=0$ and $score_x(x)=|X_x^{\prec}|\geq 1$ since $X_x^{\prec}$ contains at least $y$. We therefore have $score_z(x)>score_z(y)$;
\item [(b)] when $z\neq y$, $X$ is divisible with respect to $z$, and $z\neq x$, 
\begin{itemize}
\item [--] if $y\prec_M z$ then $y$ must be included in the bottom block of the $z$-partition and thus $score_z(y)=0$. Since the value 0 is the minimal score value that an alternative can possibly have under the DSR-scoring method, we thus have $score_z(x)\geq score_z(y)$ in the context of $y\prec_M z$; 
\item [--] else if $y\sim_M z$, as already shown in the case of $y\sim_M z$ we will have $x\succeq_M z$. Thus both $x$ and $y$ are included in the top block of the $z$-partition. On one hand, when $X_z^{\prec}=\emptyset$, the $z$-partition is a bipartition, so that $score_z(x)=\mu(x,z)$ and $score_z(y)=\mu(y,z)$. Since we already obtained $y\sim_M z$ and $x\succeq_M z$, by expression (5) we have\footnote{If the point to be assigned to both sides of a tie changes from 0.5 to another value in $[0,1]$, then the result remains unchanged.} $score_z(x)=\mu(x,z)\geq \mu(y,z)=score_z(y)$. On the other hand, when $X_z^{\prec}\neq\emptyset$, the $z$-partition can be a bipartition or a tripartition. Whatever it is, since both $x$ and $y$ are in the top block, so in both cases the formulas for computing their scores take the same form, that is, $score_z(x)=\mu(x,z)+\sum_{w\in X_z^{\prec}}\mu(x,w)$ and $score_z(y)=\mu(y,z)+\sum_{w\in X_z^{\prec}}\mu(y,w)$. Considering that (i) since $y\sim_M z$ and $x\succeq_M z$, we have $\mu(x,z)\geq \mu(y,z)$; (ii) for $w\in X_z^{\prec}$, if $y\sim_M w$ then $x\succeq_M w$ and thus $\mu(x,w)\geq\mu(y,w)$; (iii) for $w\in X_z^{\prec}$, if $y\succ_M w$ then $x\succ_M w$ and thus $\mu(x,w)=\mu(y,w)=1$; and (iv) for $w\in X_z^{\prec}$, if $y\prec_M w$ then $\mu(y,w)=0$ and thus $\mu(x,w)\geq\mu(y,w)=0$. Therefore, we have $score_z(x)=\mu(x,z)+\sum_{w\in X_z^{\prec}}\mu(x,w)\geq \mu(y,z)+\sum_{w\in X_z^{\prec}}\mu(y,w)=score_z(y)$ in the context of $y\sim_M z$ and $X_z^{\prec}\neq\emptyset$. In conjunction with our deduction under conditions of $y\sim_M z$ and $X_z^{\prec}=\emptyset$, it follows that when $y\sim_M z$ we have $score_z(x)\geq score_z(y)$;
\item [--] else (i.e., the instance of $y\succ_M z$), as already shown, when $y\succ_M z$ we will have $x\succ_M z$ hence $\mu(x,z)=1$ and $\mu(y,z)=1$. Since our hypothesis is that $X$ is divisible with respect to $z$, thus both $x$ and $y$ are included in the top block of the $z$-partition.

On one hand, when $X_z^{\prec}=\emptyset$, the $z$-partition is a bipartition, so that $score_z(x)=\mu(x,z)=1=\mu(y,z)=score_z(y)$. On the other hand, when $X_z^{\prec}\neq\emptyset$, the $z$-partition can be a bipartition or a tripartition. Whatever it is, since both $x$ and $y$ are in the top block, so in both cases the formulas for computing their scores take the same form (i.e., expression (7)), that is, $score_z(y)=\mu(y,z)+\sum_{w\in X_z^{\prec}}\mu(y,w)=1+\sum_{w\in X_z^{\prec}}\mu(y,w)$ and $score_z(x)=\mu(x,z)+\sum_{w\in X_z^{\prec}}\mu(x,w)=1+\sum_{w\in X_z^{\prec}}\mu(x,w)$. 

For every $w\in X_z^{\prec}$, we distinguish three cases: (i) when $y\succ_M w$, as already asserted in the third paragraph in this proof, we also have $x\succ_M w$. Thus we have $\mu(x,w)=\mu(y,w)=1$ in a situation where $y\succ_M w$; (ii) when $y\sim_M w$, as already asserted in the third paragraph in this proof, we have $x\succeq_M w$. Thus we have $\mu(x,w)\geq\mu(y,w)$ in a situation where $y\sim_M w$; and (iii) when $y\prec_M w$, we will have $\mu(y,w)=0$, so that we have $\mu(x,w)\geq \mu(y,w)$ in the situation $x\prec_M w$.

Therefore, we have $score_z(x)=\mu(x,z)+\sum_{w\in X_z^{\prec}}\mu(x,w)\geq \mu(y,z)+\sum_{w\in X_z^{\prec}}\mu(y,w)=score_z(y)$ when [$z\neq y$, $X$ is divisible with respect to $z$, $z\neq x$, and $y\succ_M z$]. 
\end{itemize}
Combining the last three steps, we obtain $score_z(x)\geq score_z(y)$ under the conditions $z\in X\setminus\{x,y\}$ and $X$ is divisible with respect to $z$.
\end{itemize}
Combining the results obtained in situations of $z\in X\setminus\{x,y\}$ and $z=x$, we obtain $score_z(x)\geq score_z(y)$ under the conditions $z\in X\setminus\{y\}$ and $X$ is divisible with respect to $z$.
\end{itemize}
Combining the results obtained in the situation of $z\in X\setminus\{y\}$, we obtain $score_z(x)\geq score_z(y)$ no matter whether or not $X$ is divisible with respect to $z$.

As shown above, we have obtained that $score_y(x)>score_y(y)$ and $score_z(x)\geq score_z(y)$ for $z\in X\setminus\{y\}$. It follows that we obtain $\psi(x)=\sum_{z\in X}score_z(x)>\sum_{z\in X}score_z(y)=\psi(y)$. That is, our statement regarding weak Pareto holds.

\item [(2)] \textbf{(SP)} Under the condition that the individuals' preferences are orderings, we need to show that $\forall x,y \in X$, if $\forall k(x\succeq_k y)$ and $\exists k(x\succ_k y)$ then $\psi(x)>\psi(y)$. The reasoning is quite similar to that when to assert our statement regarding the weak Pareto since in fact, from $[(\forall k(x\succeq_k y))\land(\exists k(x\succ_k y))]$ and [each individual's preference is an ordering of the alternatives] we can deduce two results similar to those that played a fundamental role in the previous proving process. That is, $\forall z(y\succ_M z\rightarrow x\succ_M z)$ and $\forall z(y\sim_M z\rightarrow x\succeq_M z)$. Of course, $x\succ_M y$ holds as well under the simple majority rule. Similar analysis will lead to a summary which will assert that our statement regarding the strong Pareto is valid.
\item [(3)] \textbf{(ND)} The property of non-dictatorship (ND) trivially holds for the DSR-scoring method since, apparently, the expression (7) or (9) yields a social ranking that does not always coincide fully with the preference of a certain individual.
\item [(4)] \textbf{(IIA)} An easy proof of this assertion can be directly deduced from Lemma 1. We introduce some notations to be used in this proof.
\begin{itemize}
\item [--] Let $\{x,y\}$ be a subset of $X$. 
\item [--] Suppose that $\varrho=(O_1,O_2,\dots,O_n)$ and $\varrho'=(O_1',O_2',\dots,O_n')$ are two profiles such that $\varrho|\{x,y\}=\varrho'|\{x,y\}$.
\item [--] Let $M_{\varrho}$ and $M_{\varrho'}$ denote the majority relations derived respectively from $\varrho$ and $\varrho'$. 
\item [--] Let $\psi_{\varrho}(x)$ and $\psi_{\varrho}(y)$ represent the DSR-scores of $x$ and $y$, respectively, on the basis of $\varrho$. 
\item [--] Let $\psi_{\varrho'}(x)$ and $\psi_{\varrho'}(y)$ represent the DSR-scores of $x$ and $y$, respectively, on the basis of $\varrho'$.
\end{itemize}
We need to assert that if both $M_{\varrho}$ and $M_{\varrho'}$ are orderings on $X$, then we will obtain: $\psi_{\varrho}(x)>\psi_{\varrho}(y)$ iff $\psi_{\varrho'}(x)>\psi_{\varrho'}(y)$; $\psi_{\varrho}(x)=\psi_{\varrho}(y)$ iff $\psi_{\varrho'}(x)=\psi_{\varrho'}(y)$; and $\psi_{\varrho}(x)<\psi_{\varrho}(y)$ iff $\psi_{\varrho'}(x)<\psi_{\varrho'}(y)$.

Because of $\varrho|\{x,y\}=\varrho'|\{x,y\}$, thus we have $x M_{\varrho} y=x M_{\varrho'} y$ in the sense that
\begin{itemize}
\item [(a)] $x \succ_{M_{\varrho}} y$ iff $x \succ_{M_{\varrho'}} y$;
\item [(b)] $x \sim_{M_{\varrho}} y$ iff $x \sim_{M_{\varrho'}} y$;
\item [(b)] $x \prec_{M_{\varrho}} y$ iff $x \prec_{M_{\varrho'}} y$.
\end{itemize}
Since our hypothesis is that both $M_{\varrho}$ and $M_{\varrho'}$ are orderings on $X$, by Lemma 1, we know that
\begin{itemize}
\item [(d)] if $x \succ_{M_{\varrho}} y$ then $\psi_{\varrho}(x)>\psi_{\varrho}(y)$; if $x \succ_{M_{\varrho'}} y$ then $\psi_{\varrho'}(x)>\psi_{\varrho'}(y)$;
\item [(e)] if $x \sim_{M_{\varrho}} y$ then $\psi_{\varrho}(x)=\psi_{\varrho}(y)$; if $x \sim_{M_{\varrho'}} y$ then $\psi_{\varrho'}(x)=\psi_{\varrho'}(y)$;
\item [(f)] if $x \prec_{M_{\varrho}} y$ then $\psi_{\varrho}(x)<\psi_{\varrho}(y)$; if $x \prec_{M_{\varrho'}} y$ then $\psi_{\varrho'}(x)<\psi_{\varrho'}(y)$.
\end{itemize}
Combining the steps (a)-(f), we obtain what we wish to assert, namely,
\begin{itemize}
\item [(i)] $\psi_{\varrho}(x)>\psi_{\varrho}(y)$ iff $\psi_{\varrho'}(x)>\psi_{\varrho'}(y)$,
\item [(ii)] $\psi_{\varrho}(x)=\psi_{\varrho}(y)$ iff $\psi_{\varrho'}(x)=\psi_{\varrho'}(y)$,
\item [(iii)] $\psi_{\varrho}(x)<\psi_{\varrho}(y)$ iff $\psi_{\varrho'}(x)<\psi_{\varrho'}(y)$,
\end{itemize}
so that our statement regarding IIA is valid. Q.E.D.
\end{itemize}

\vspace{0.2cm}

In addition, the properties of anonymity and neutrality hold trivially for the DSR-scoring method when it is used as a SWF as a result of the fact that: (1) it treats all individuals in an equal way; and (2) it treats all alternatives in the same way.

\vspace{0.2cm}

As it is well known, single-peaked domain (e.g., the number of individuals is odd, and the individual preferences are linear and single-peaked on a line) guarantees the transitivity of pairwise majority relation (Black, 1948; Arrow, 1951; Puppe, 2018)\footnote{Single-peakedness is not the only domain restriction allowing for a transitive SWF. Many successful restrictive assumptions on individual preferences were reported including, among others, the single-cavedness (Inada, 1964), the condition of dichotomous preferences (Inada, 1964), the Latin-square-lessness (Ward, 1965), and the value restriction (Sen, 1966, 1970). In addition to these sufficient conditions, necessary and sufficient conditions in the present context were also observed by Sen and Pattanaik (1969). For detailed investigation of various forms of domain restrictions for different SWFs/SCFs, the reader is referred to, e.g., Sen (1970), Pattanaik (1970), Gaertner (2002), and Dietrich and List (2010).}. In conjunction with Lemma' and the IIA statement of Theorem 3, we conclude that, \textit{if the preference profiles are restricted to a single-peaked domain, then we know that: (1)the DSR-scoring method satisfies Arrow conditions of Social transitivity, WP, SP, IIA and ND; (2) the collective preference relation determined by the DSR-scoring method will coincide with that of majority relation}.

\vspace{0.2cm}

\textbf{Theorem 4} When the DSR-scoring method is used as a SWF/SCC, it has the following properties:
\begin{itemize}
\item [(1)] \textbf{(Condorcet winner principle)} As a SWF/SCC, the DSR-scoring method is Condorcet consistent\footnote{A SCC is Condorcet consistent if it always selects the Condorcet winner as the uniquely optimal choice whenever one exists (Brandt et al., 2022)}.
\item [(2)] \textbf{(Condorcet loser principle)} As a SWF/SCC, the DSR-scoring method satisfies the Condorcet loser principle.
\item [(3)] \textbf{(Strong Gehrlein-stability)} As a SWF/SCC, the DSR-scoring method is Strongly Gehrlein-stable. Formally, if $A\succ B$, where $A,B\in 2_+^X$, $A\cap B\neq\emptyset$ and $A\cup B=X$, then $\forall a\in A\forall b\in B:\psi(a)>\psi(b)$.
\end{itemize}

\textbf{Proof}\begin{itemize}
\item [(1)] (Condorcet winner principle) We need only to show that, if $x$ is a Condorcet winner (CW), then it will be assigned a unique maximal score by the DSR-scoring method. Formally, we need to prove: \par
if $x\succ_M y, \forall y\in X\setminus \{x\}$, and if we use the DSR-scoring method (expression (9)) for computing the alternatives' scores, then we will have $\psi(x)>\psi(y), \forall y\in X\setminus \{x\}$.\par
Expression (9) formulizes the scoring strategy introduced in subsection 4.1. Under our strategy, a certain alternative's final score is the sum of those scores derived from checking each alternative's possible specific partition: a certain alternative $x$'s final score is denoted by $\psi(x)=\sum_{z\in X}score_z(x)$, where $score_z(x)$ can be a nonzero value only when the alternative set $X$ is divisible with respect to $z$ and $x$ is not contained in the bottom block of the alternative-specific DSRP with respect to $z$. Moreover, if $score_z(x)=\mu(x,z)+\sum_{w\in X_z^{\prec}}\mu(x,w)$ has a positive value then it will not exceed $1+|X_z^{\prec}|$. Note that we assume $|X|>1$ in this paper. We have
\begin{itemize}
\item [--] The alternative set $X$ is divisible with respect to $x$, and the candidate-specific DSRP is $\langle \{x\},X\setminus \{x\}\rangle_{x}$ since $\{x\}\succ (X\setminus \{x\})$\footnote{Note that here the introduced `specific bipartition' defined by Def.4(2) arises as an indispensable part in our `designing' a SWF/SCC.}. We therefore know that $score_x(x)>score_x(y), \forall y\in X\setminus \{x\}$, since $score_x(x)=\mu(x,x)+\sum_{w\in X_x^{\prec}}\mu(x,w)=|X|-1=m-1$ and $score_x(y)=0, \forall y\in X\setminus \{x\}$ as a result of the fact that $X\setminus \{x\}$ is the bottom block in the $x$-partition.
\item [--] For $z\in X\setminus \{x\}$, $X$ is divisible with respect to $z$ (note that, in this step, our hypothesis is $z\neq x$), then $score_z(x)\geq score_z(y), \forall y\in X\setminus\{x\}$ since: (a) when $X_z^{\prec}=\emptyset$, the $z$-partition is $\langle X\setminus \{z\},\{z\}\rangle_{z}$, where $X\setminus \{z\}=X_z^{\succeq,\neq}$. We first know that $score_z(z)=0$ because $z$ is included in the bottom block. Second, the alternative $x$ must be included in the top block because $x$ is a CW hence $x\succ_M z$ and thus $score_z(x)=\mu(x,z)=1$. Finally, because $\langle X\setminus \{z\},\{z\}\rangle_{z}$ is a bipartition with respect to $z$ under the condition that $X_z^{\prec}=\emptyset$, thus $score_z(y)=\mu(y,z)\leq 1, \forall y\in X\setminus \{z\}$. We therefore have $score_z(x)=1\geq score_z(y), \forall y\in X\setminus\{x\}$ when $X_z^{\prec}=\emptyset$; and (b) when $X_z^{\prec}\neq\emptyset$, since $x$ is a CW, we thus know $x\in X_z^{\succeq,\neq}$ and $\{x\}\succ X_z^{\prec}$. Hence by Def.4(3) we know that the $z$-partition can be\footnote{Note that here the introduced `specific bipartition' defined by Def.4(3) arises as an indispensable part in our `designing' a SWF/SCC.} $\langle X_z^{\succeq,\neq}, \{z\},X_z^{\prec}\rangle_{z}$. The alternative $x$ must be included in the top block because we have $\{x\}\succ X\setminus \{x\}$ and $z\in X\setminus \{x\}$. Moreover, because $\{x\}\succ X\setminus \{x\}$ and thus $\{x\}\succ (\{z\}\cup X_z^{\prec})$, we therefore have $score_z(x)=1+|X_z^{\prec}|\geq score_z(y), \forall y\in X\setminus\{x\}$, since no matter which block of $\langle X_z^{\succeq,\neq}, \{z\},X_z^{\prec}\rangle_{z}$ $y$ belongs to, the value of $score_z(y)$ will never exceed $1+|X_z^{\prec}|$.
\end{itemize}
Combining the last two steps, we get that $\psi(x)=\sum_{z\in X}score_z(x)>\sum_{z\in X}score_z(y)=\psi(y), \forall y\in X\setminus \{x\}$. 
The statement (1) of the theorem is proved.
\item [(2)] (Condorcet loser principle) Suppose that $x$ is a Condorcet loser (CL). Then the proof will be to assert that for any $y\in X\setminus \{x\}$ we have $\psi(x)<\psi(y)$.

We will prove the statement from three respects. First, we will show that the alternative set is divisible with respect to $x$ resulting in $score_x(x)=0$ and $\forall y(y\in X\setminus\{x\}\rightarrow score_x(y)=1)$. Second, no matter whether or not $X$ is divisible with respect to $z$ with $z\neq x$, we will prove that $\forall y(y\in X\setminus\{x\}\rightarrow score_z(y)\geq score_z(x)$. Finally, the statement follows very easily from expression (9).
\begin{itemize}
\item [(a)] Since $x$ is a CL, i.e.,  $\forall y(y\in X\setminus\{x\}\rightarrow y\succ_M x)$, so we have $X\setminus\{x\}\succ\{x\}$. By Def.4 we know that $X$ is specifically divisible with respect to $x$ and the $x$-partition is $\langle X\setminus \{x\},\{x\}\rangle_{x}$. As $x$ is in the bottom block, by expression (7), we have $score_x(x)=0$. For any $y\in X\setminus\{x\}$, since $y$ is included in the top block, and our hypothesis is $\forall y(y\in X\setminus\{x\}\rightarrow y\succ_M x)$, also by expression (7), we have $\forall y(y\in X\setminus\{x\}\rightarrow score_x(y)=\mu(y,x)=1)$. The summation in this step is that $score_x(y)> score_x(x)$ for any $y\in X\setminus\{x\}$.
\item [(b)] For any $z\in X\setminus\{x\}$, 
\begin{itemize}
\item [(i)] if $X$ is NOT divisible with respect to $z$, by expression (7) we have $score_z(w)=0$ for all $w\in X$.
\item [(ii)] else (i.e., $X$ is divisible with respect to $z$), since $x$ is a CL and $z\neq x$, we thus have $z\succ_M x$. Therefore, $x$ must be included in the bottom block of a $z$-partition given that $X$ is divisible with respect to $z$, so that we have $score_z(x)=0$. As the value 0 is the minimal possible value of $score_z(y)$ for all $y$, we therefore have $score_z(y)\geq score_z(x)$ for all $y\in X$.
\end{itemize}

\item [] The summation in this step is that $score_z(y)\geq score_z(x)$ for any $y,z\in X\setminus\{x\}$.
\item [(c)] Combining the results of the above two steps (a)-(b), and by expression (9), we know $\psi(y)=score_x(y)+\sum_{z\in X\setminus\{x\}}score_z(y)> score_x(x)+\sum_{z\in X\setminus\{x\}}score_z(x)=\psi(x)$. The statement (2) of the theorem is proved.
\end{itemize}
 
\item [(3)] (Strong Gehrlein-stability) Our hypotheses are: $|X|>1$, $A,B\in 2^X_+$, $A\cup B=X$,  $A\cap B=\emptyset$ (hence $|A|\geq 1$ and $|B|\geq 1$), and $A\succ B$.

We need to prove: $\forall a\in A\forall b\in B:\psi(a)>\psi(b)$.

As $A,B\in 2_+^X, A\cup B=X$, and $A\cap B=\emptyset$, by expression (9), we know
\begin{itemize}
\item [\checkmark] $\psi(a)=\sum_{z\in X}score_z(a)=\sum_{x\in A}score_x(a)+\sum_{y\in B}score_y(a)$; and
\item [\checkmark] $\psi(b)=\sum_{z\in X}score_z(b)=\sum_{x\in A}score_x(b)+\sum_{y\in B}score_y(b)$.
\end{itemize}
Under our hypotheses, we will show that $\sum_{x\in A}score_x(a)>0=\sum_{x\in A}score_x(b)$, and $\sum_{y\in B}score_y(a)\geq \sum_{y\in B}score_y(b)$, so that $\psi(a)>\psi(b)$ will follow easily.
\begin{itemize}
\item [(I)] For any $a\in A$, we must have $X_a^{\prec}\neq\emptyset$ and $B\subseteq X_a^{\prec}$, since $a\in A$, $A\succ B$ and $|B|\geq 1$. The alternative set $X$ is divisible with respect to $a$ as a result of the fact that: 
\begin{itemize}
\item [(i)] if $X_a^{\succeq,\neq}\neq\emptyset$ (that is, if there exists alternative $z$ such that $z\succeq_M a$ with $z\neq a$), then those alternatives in $X_a^{\succeq,\neq}$ must come from $A$, namely, $X_a^{\succeq,\neq}\subset A$, since we are given $A\succ B$ (in words, every alternative in $B$ is beaten by each of $A$). Recalling that $\gamma(S_1,S_2)>0$ is interpreted as `each alternative in $S_1$ beats some alternative in $S_2$', we have $\gamma(X_a^{\succeq,\neq}\cup\{a\},X_a^{\prec})>0$, since we are given $a\in A$ and $A\succ B$, and, we just obtained $X_a^{\succeq,\neq}\subset A$ and $B\subseteq X_a^{\prec}$. Besides, recalling that $\beta(S_1,S_2)>0$ is interpreted as `there exists some alternative in $S_1$ that beats all the alternatives in $S_2$', we know that $\beta(X_a^{\succeq,\neq}\cup\{a\},X_a^{\prec})>0$ is trivially satisfied since $\{a\}\succ X_a^{\prec}$. By Def.4 we know that $X$ is divisible with respect to $a$ provided that $X_a^{\succeq,\neq}\neq\emptyset$. Particularly, in this situation the $a$-partition\footnote{Note that, here the `specific bipartition' of the second-type defined by Def.2 and Def.4(2) arises as an integral part in our `designing' a SWF/SCC with strong Gehrlein stability.} is $\langle X_a^{\succeq,\neq}\cup\{a\},X_a^{\prec}\rangle_{a}$.
\item [(ii)] else (i.e., $X_a^{\succeq,\neq}=\emptyset$), by Def.4, we know that $X$ is divisible with respect to $a$, since $\{a\}\succ X_a^{\prec}$ implies both $\{a\}\succeq X_a^{\prec}$ and $\beta(\{a\}, X_a^{\prec})>0$. Particularly, in this situation the $a$-partition is $\langle \{a\},X_a^{\prec}\rangle_{a}$.
\end{itemize}
According to expression (9), we know that $score_a(a)>0$ since $a$ is contained in the top block either of $\langle X_a^{\succeq,\neq}\cup\{a\},X_a^{\prec}\rangle_{a}$ or $\langle \{a\},X_a^{\prec}\rangle_{a}$, whereas for all $b\in B$ we have $score_a(b)=0$ since $B$ is included in the bottom block $X_a^{\prec}$. Due to $a$ being arbitrarily selected from $A$, we further have $\forall x\in A \forall b\in B(score_x(b)=0)$. In addition, as repeatedly mentioned in previous proofs, the minimum value of `$score$' is 0. We thus have $score_x(a)\geq 0, \forall x\in A$. Therefore, it follows that
$$\sum_{x\in A}score_x(a)=score_a(a)+\sum_{x\in A\setminus\{a\}}score_x(a)>0=\sum_{x\in A}score_x(b), \forall a\in A\forall b\in B.\eqno(*)$$
\item [(II)] Similarly, for any $b\in B$, we must have $X_b^{\succeq, \neq}\neq\emptyset$ and $A\subseteq X_b^{\succeq, \neq}$, since $b\in B$ and $A\succ B$. The alternative set $X$ is divisible with respect to $b$ as a result of the fact that: 
\begin{itemize}
\item [(i)] if $X_b^{\prec}\neq\emptyset$, that is, if there exists alternative $z$ such that $z\prec_M b$ with $z\neq b$, then those alternatives in $X_b^{\prec}$ must come from $B$, namely, $X_b^{\prec}\subset B$, since we are given $A\succ B$ and $b\in B$. Given $X_b^{\prec}\subset B$, we have $A\succ X_b^{\prec}$ since $A\succ B$. We thus know $\beta(X_b^{\succeq,\neq}\, X_b^{\prec})>0$ since we just obtained $A\subseteq X_b^{\succeq, \neq}$ and $A\succ X_b^{\prec}$. By Def.4 we know that $X$ is divisible with respect to $b$ provided that $X_b^{\prec}\neq\emptyset$. Particularly, in this situation the $b$-partition\footnote{Note that, here the `specific tripartition' defined by Def.4(3) arises as an integral part in our `designing' a SWF/SCC with strong Gehrlein stability.} is $\langle X_b^{\succeq,\neq}, \{b\}, X_b^{\prec}\rangle_{b}$.
\item [(ii)] else (i.e., $X_b^{\prec}=\emptyset$), by Def.4, we know that $X$ is divisible with respect to $b$, mainly because [$A\subset X_b^{\succ,\neq}$, $A\succ B$, $b\in B$ and $|A|\geq 1$] result in $\beta(X_b^{\succ, \neq}, \{b\})>0$. Particularly, in this situation the $b$-partition is $\langle X_b^{\succ, \neq}, \{b\}\rangle_{b}$.
\end{itemize}
According to expression (9), since $A$ is included in the top block either of $\langle X_b^{\succeq,\neq}, \{b\}, X_b^{\prec}\rangle_{b}$ or $\langle X_b^{\succ, \neq}, \{b\}\rangle_{b}$, and $A\succ X_b^{\prec}$ as well as $A\succ \{b\}$, we know that the alternatives in $A$ all take the maximum `$score$' value of $score_b(x)=\mu(x,b)+\sum_{z\in X_b^{\prec}}(x,z)=1+|X_b^{\prec}|=\max_{w\in X}score_b(w), \forall x\in A$. We thus have $score_b(x)\geq score_b(y), \forall x\in A\forall y\in B$. Due to $b$ being arbitrarily selected from $B$, it follows that
$$\sum_{y\in B}score_y(a)\geq \sum_{y\in B}score_y(b), \forall a\in A\forall b\in B.\eqno(**)$$
\end{itemize}
Combining the above two inequalities (*) and (**), we obtain
$$\psi(a)=\sum_{x\in A}score_x(a)+\sum_{y\in B}score_y(a)>\sum_{x\in A}score_x(b)+\sum_{y\in B}score_y(b)=\psi(b), \forall a\in A\forall b\in B,$$
so that the statement (3) of the theorem is proved. Q.E.D.
\end{itemize}

\vspace{2mm}

\textbf{Remark 2} The reader may have observed from the role of scores relavant to `$\sim$' in the above proof procedures that, if the value of $\mu(x\sim y)$ is reset as a value $\alpha$ from the interval $[0,1]$ rather than 0.5 provided that $\mu(x\succ y)=1$ and $\mu(x\prec y)=0$ as shown by expression (5), then our statements in Theorems 3 and 4 will not be altered. By the way, since strong Gehrlein-stability implies Smith set principle, therefore, we know that, when the DSR-scoring method is used as a SWF/SCC, it satisfies the Smith set principle.

\vspace{2mm}

The statement (1) of Theorem 4 indicates that the DSR-scoring method, when used as a SWF/SCC, is Condorcet consistent. According to ``\textit{Condorcet's principle implies the no-show paradox}''(Moulin 1988)\footnote{The no-show paradox refers to a phenomenon that a voter may be better off by not participating in the election (Fishburn and Brams, 1983). For a more recent exploration, the reader may consult Brandt et al. (2021).}, we state the following result without proof.

\textbf{Proposition 4} When used as a SWF/SCC, the DSR-scoring method does not guarantee the exclusion of no-show paradox.

\vspace{2mm}

We note that, the statement (1) of Theorem 4 can be regarded as a corollary of the statement (3), since the dominating set can be regarded as an extension of the CW.

%


\vspace{0.2cm}

As an illustration for the property of Pareto, we reconsider Example 2 (as aforementioned, taken from Gaertner (2009, Page 110)), where alternative $u$ is considered by all voters as less preferred than candidate $z$ (Gaertner, 2009, Page 110). Therefore, if $z$ is selected then $u$ should not. This desired result will be guaranteed by the DSR-scoring method as shown by Table 4, from which we obtain the alternatives' DSR-scores as follows 
$$\begin{array}{cc}
\psi(x)=\sum_{a\in X}score_a(x)=0,& \psi(y)=\sum_{a\in X}score_a(y)=3,\\
\psi(z)=\sum_{a\in X}score_a(z)=4,& \psi(u)=\sum_{a\in X}score_a(u)=1,
\end{array}$$
so that the social ranking is $z\succ y\succ u \succ x$.

\begin{table}[h]
\renewcommand\arraystretch{1.3}
\centering\footnotesize
\caption{Scoring process of Example 2}
\begin{tabular}{ccccccc}
    \toprule
    \multicolumn{1}{c}{$x\succ y, z\succ x, u\succ x,$}  & \multicolumn{4}{c}{alternative ($a$)} \\
    \cmidrule{2-3}  \cmidrule{4-5}  \cmidrule{6-7}
    {$y\succ z, y\succ u, z\succ u$}  & $x$ & $y$ & $z$ & $u$ \\
    \midrule
$X_a^{\succeq,\neq}$ &

$\{z,u\}
$ &
$\{x\}
$& $\{y\}$ & $\{y,z\}$ \\

$\{a\}$ &

$\{x\}
$ &
$\{y\}
$& $\{z\}$ & $\{u\}$ \\

$X_a^{\prec}$ &

$\{y\}
$ &
$\{z,u\}
$& $\{x,u\}$ & $\{x\}$ \\

$a$-partition &

{-} & {-} & {$\left \langle {\begin{array}{cc} {\{y,z\}}\\
{\{x,u\}}
\end{array}}\right \rangle$$_z$} & {$\left \langle {\begin{array}{cc} {\{y,z\}}\\
{\{u\}}\\
{\{x\}}
\end{array}}\right \rangle$$_u$} \\

$score_a(x)$ &

{0} & {0} & {0} & {0} \\

$score_a(y)$ &

{0} & {0} & {2} & {1} \\

$score_a(z)$ &

{0} & {0} & {2} & {2} \\

$score_a(u)$ &

{0} & {0} & {0} & {1} \\

    \bottomrule
\end{tabular}
\end{table}

\subsection{Using a tournament matrix as input}

Since the input data of the DSR-scoring method are required to be complete binary preference relations over a finite alternative set (without the prerequisite of transitivity), considering that the tournament is a complete binary relation over a finite alternative set (without ties), thus the DSR-scoring method can use a tournament matrix as input data and then output the tournament solution (i.e., the choice set of the tournament\footnote{See Laslier (1997) for the formal definition of tournament solution.}). Let $T(X)$ denote a tournament on $X$, and $\psi_{T(X)}(x)$ the DSR-score of alternative $x$. We also use $T(X)$ to denote the tournament matrix. When the DSR-scoring method is employed to seek the tournament solution based on a tournament matrix, the tournament solution is formulized as
$$DSR(T(X))=\arg\max_{x\in X}\psi_{T(X)}(x).$$

To illustrate, we consider the following example (Dutta, 1988). 

\textbf{Example 3} The alternative set is $\{a_1,a_2,a_3,a_4,a_5,a_6\}$. The dominance relation and the tournament matrix are included in Figure 3, where `$x\rightarrow y$' indicates `$x\succ y$'. 

\begin{figure}[H]
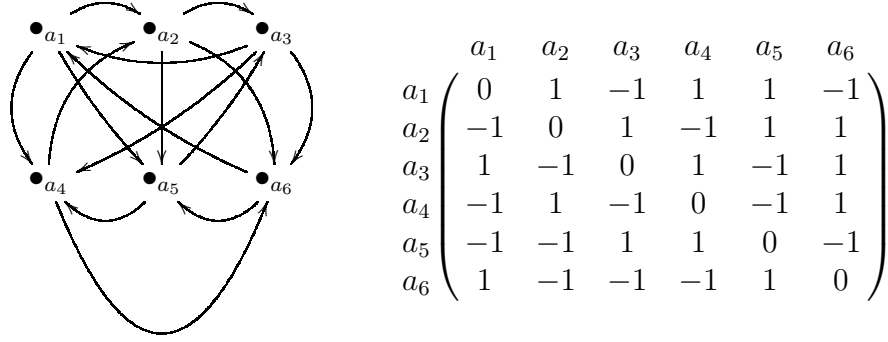

\centering
\subfigure{}
{
\begin{minipage}{5cm}
\[ \xygraph{  
        !{<0cm,0cm>;<1cm,0cm>:<0cm,1cm>::}  
        !{(0,0) }*+{\bullet_{a_4}}="a4"  
        !{(0,2) }*+{\bullet_{a_1}}="a1"  
        !{(1.5,2) }*+{\bullet_{a_2}}="a2"  
        !{(3,2) }*+{\bullet_{a_3}}="a3" 
        !{(3,0)}*+{\bullet_{a_6}}="a6"  
        !{(1.5,0)}*+{\bullet_{a_5}}="a5"
            "a1":@/^0.4cm/"a2"^(0.6){}  
            "a1":@/_0.5cm/"a4"^(0.6){}
            "a1":@/_0.1cm/"a5" ^(0.8){}   
            "a2":@/^0.4cm/"a3" ^(0.8){}  
            "a2":@/^0cm/"a5" ^(0.7){} 
            "a2":@/^0.5cm/"a6" ^(0.7){}      
             "a3":@/^0.4cm/"a1" ^(0.8){}  
            "a3":@/^/"a4" ^(0.4){}  
            "a3":@/^0.5cm/"a6" ^(0.7){}   
            "a4":@/^0.5cm/"a2" ^(0.4){}  
            "a4":@/_2cm/"a6" ^(0.7){}
            "a5":@/_0.1cm/"a3" ^(0.4){}  
            "a5":@/^0.5cm/"a4" ^(0.7){}
            "a6":@/^/"a1" ^(0.4){}  
            "a6":@/^0.5cm/"a5" ^(0.7){}
        }
\]  
\end{minipage}
}
\subfigure{}
{
\begin{minipage}{5cm}
\[
\bordermatrix{
    & a_1 & a_2 & a_3 & a_4 & a_5 & a_6\cr
  a_1 & 0 & 1 & -1 & 1 & 1 & -1\cr
  a_2 & -1 & 0 & 1 & -1 & 1 & 1\cr
  a_3 & 1 & -1 & 0 & 1 & -1 & 1\cr
  a_4 & -1 & 1 & -1 & 0 & -1 & 1\cr
  a_5 & -1 & -1 & 1 & 1 & 0 & -1\cr
  a_6 & 1 & -1 & -1 & -1 & 1 & 0
}
\]
\end{minipage}
}
\caption{Tournament diagram and tournament matrix in Example 3.}
\end{figure}

\begin{table}[h]
\renewcommand\arraystretch{1.3}
\centering\footnotesize
\caption{Analysis process of Example 3}
\begin{tabular}{ccccccc}
    \toprule
    \multicolumn{1}{c}{}  & \multicolumn{6}{c}{alternative ($x$)} \\
    \cmidrule{2-3}  \cmidrule{4-5}  \cmidrule{6-7}
    {}  & $a_1$ & $a_2$ & $a_3$ & $a_4$ & $a_5$ & $a_6$\\
    \midrule
$X_x^{\succeq,\neq}$ &

$\{a_3,a_6\}
$ &
$\{a_1,a_4\}
$& $\{a_2,a_5\}$ & $\{a_1,a_3,a_5\}$ & $\{a_1,a_2,a_6\}$ & $\{a_2,a_3,a_4\}$\\

$\{x\}$ &

$\{a_1\}
$ &
$\{a_2\}
$& $\{a_3\}$ & $\{a_4\}$ & $\{a_5\}$ & $\{a_6\}$\\

$X_x^{\prec}$ &

$\{a_2,a_4,a_5\}
$ &
$\{a_3,a_5,a_6\}
$& $\{a_1,a_4,a_6\}$ & $\{a_2,a_6\}$ & $\{a_3,a_4\}$ & $\{a_1,a_5\}$\\

$x$-partition &

{$\left \langle {\begin{array}{cc} {\{a_1,a_3,a_6\}}\\
{\{a_2,a_4,a_5\}}
\end{array}}\right \rangle$$_{a_1}$} & {$\left \langle {\begin{array}{cc} {\{a_1,a_2,a_4\}}\\
{\{a_3,a_5,a_6\}}
\end{array}}\right \rangle$$_{a_2}$} & {$\left \langle {\begin{array}{cc} {\{a_2,a_3,a_5\}}\\
{\{a_1,a_4,a_6\}}
\end{array}}\right \rangle$$_{a_3}$} & {-} & {-} & {-} \\

$score_x(a_1)$ &

{3} & {2} & {0} & {0} & {0} & {0} \\

$score_x(a_2)$ &

{0} & {3} & {2} & {0} & {0} & {0} \\

$score_x(a_3)$ &

{2} & {0} & {3} & {0} & {0} & {0} \\

$score_x(a_4)$ &

{0} & {2} & {0} & {0} & {0} & {0} \\

$score_x(a_5)$ &

{0} & {0} & {2} & {0} & {0} & {0} \\

$score_x(a_6)$ &

{2} & {0} & {0} & {0} & {0} & {0} \\

    \bottomrule
\end{tabular}
\end{table}

The analysis process is summarized in Table 5. In this example, the alternative set is divisible with respect to $a_1$, $a_2$ and $a_3$. The three alternative-specific DSRPs belong to the second-type bipartition, which can be easily checked by Def.2 and Def.4. By expression (6), we obtain the DSR-scores of alternatives:
$$\begin{array}{cccc}
\psi(a_1)=\sum_{x\in X}score_x(a_1)=5,& \psi(a_2)=\sum_{x\in X}score_x(a_2)=5, & \psi(a_3)=\sum_{x\in X}score_x(a_3)=5,\\
\psi(a_4)=\sum_{x\in X}score_x(a_4)=2,& \psi(a_5)=\sum_{x\in X}score_x(a_5)=2, & \psi(a_6)=\sum_{x\in X}score_x(a_6)=2,
\end{array}$$
so that the ranking is $a_1\sim a_2\sim a_3\succ a_4\sim a_5\sim a_6$, and the tournament solution is $\{a_1,a_2,a_3\}$. This result coincides with Dutta's expectations (Dutta, 1988, P67), as he put it ``\textit{then $\{a_1,a_2,a_3\}$ instead of ... is an interesting possibility}''.

\subsection{Under dichotomous preferences}

Approval ballot, as a convenient expression means for individuals' dichotomous preferences, asks individuals to select whether to approve or disapprove of each alternative (Xu, 2010; Brandl and Peters, 2022). The approval voting, proposed by Brams and Fishburn (1978, 1983), elects the alternatives with the highest number of approvals. Here we are not concerned with the theoretical and empirical aspects of the approval voting. Rather, we will show that the DSR-scoring method may be used as an approach of aggregating approval ballots. In our approach, the individuals' dichotomous preferences are first transformed into weak orderings by making use of their distinctive preference structures where each approval ballot partitions the alternative set into approved alternatives and disapproved alternatives. An individual views all approved alternatives as equally important and all disapproved alternatives as equally important too, but each approved alternative as more preferred to each disapproved alternative. Then a majority relation will be derived from those weak orderings and will further be provided to the DSR-scoring method so as to get a final ranking or choice set of the alternatives. We demonstrate it on an example which is taken from Diss et al. (2010).

\textbf{Example 4} Eleven voters provide their individual preferences over $X=\{a,b,c\}$ as follows (the numbers indicate how many individuals holding the  subsequent preferences, and the alternatives with underlines are approved ones)
$$\begin{array}{ll}
 \text{4}: & \underline{a} \succ b \succ c ,\\
 \text{2}: & \underline{b} \succ a \succ c ,\\
 \text{1}: & \underline{b} \succ \underline{c} \succ a ,\\
 \text{4}: & \underline{c} \succ b \succ a.
\end{array}
$$

Under our scheme, the individuals' dichotomous preferences are transformed into weak orderings:
$$\begin{array}{ll}
 \text{4}: & a \succ b \sim c ,\\
 \text{2}: & b \succ a \sim c ,\\
 \text{1}: & b \sim c \succ a ,\\
 \text{4}: & c \succ b \sim a,
\end{array}
$$
from which we derive the following pairwise majority relations
$$a\succ b, c\succ a, c\succ b.$$

Then we use Expression (7) to compute the DSR-scores of $score_x(y)$. The analysis process is shown by Table 6.

\begin{table}[h]
\renewcommand\arraystretch{1.3}
\centering\footnotesize
\caption{Computing the DSR-scores of $score_x(y)$ in Example 4}
\begin{tabular}{ccccccc}
    \toprule
    \multicolumn{1}{c}{majority relation}  & \multicolumn{3}{c}{alternative ($x$)} &\\
    \cmidrule{2-3}  \cmidrule{4-5} \cmidrule{6-7}
    {$a\succ b, c\succ a, c\succ b$}  & $a$ & $b$ & $c$ &\\
    \midrule
$X_x^{\succeq,\neq}$ &

$\{c\}
$ &
$\{a,c\}
$& $\{\}$ &\\

$\{x\}$ &

$\{a\}
$ &
$\{b\}
$& $\{c\}$ &\\

$X_x^{\prec}$ &

$\{b\}
$ &
$\{\}
$& $\{a,b\}$ &\\

$x$-partition &

{$\left \langle {\begin{array}{cc} {\{c\}}&\\
{\{a\}}\\
{\{b\}}
\end{array}}\right \rangle$$_a$} & {$\left \langle {\begin{array}{cc} {\{a,c\}}&\\
{\{b\}}
\end{array}}\right \rangle$$_b$} & {$\left \langle {\begin{array}{cc} {\{c\}}&\\
{\{a,b\}}
\end{array}}\right \rangle$$_c$} &\\

$score_x(a)$ &

{1} & {1} & {0} \\

$score_x(b)$ &

{0} & {0} & {0} \\

$score_x(c)$ &

{2} & {1} & {2} \\

    \bottomrule
\end{tabular}
\end{table}

By expression (6), we obtain the DSR-scores of alternatives:
$$
\psi(a)=\sum_{x\in \{a,b,c\}}score_x(a)=2, \psi(b)=\sum_{x\in \{a,b,c\}}score_x(b)=0, \psi(c)=\sum_{x\in \{a,b,c\}}score_x(c)=5,
$$
so that the choice set is $\{c\}$, which coincides with the outcome of the approval voting method.

\vspace{0.1cm}

It is pointed out by Inada (1964) that, if the individual preferences are provided as dichotomous preferences, then the pairwise majority relation is transitive on the alternative set. In conjunction with the statement regarding IIA of Theorem 3 and Theorem 4, we have the following corollary.

{\bf Corollary 4} When the individual preferences are dichotomous preferences, the DSR-scoring method satisfies the conditions of Condorcet winner principle, Condorcet loser principle, strong Gehrlein-stability (hence Smith set principle), anonymity, neutrality, weak Pareto (WP), strong Pareto (SP), non-dictatorship (ND), and independence of irrelevant alternatives (IIA).

Since the pairwise majority relation on the alternative set is transitive when the individual preferences are dichotomous (Inada, 1964), from Lemma 1' we know that, the collective preference relation determined by the DSR-scoring method coincides with that of majority relation in the dichotomous case. According to ``\textit{approval voting and majority voting
happen to share the same collective preference function on the dichotomous domain}'' (Maniquet and Mongin, 2015), we further know that the DSR-scoring method shares the same collective preference function on the dichotomous domain with the approval voting and the majority voting.


\section{Comparisons and relationships to some existing methods}

\subsection{A comparison in relation to computational complexity with the Dodgson's rule, the Kemeny's rule, the Slater's rule, the Banks rule, and the Schwartz’s tournament equilibrium set (TEQ) rule}

It is well-known that the Dodgson's rule (an alternative is chosen as a Dodgson winner if it has a minimal number of adjacent pairwise interchanges in individuals' preference orders to become a Condorcet winner, see Black (1958)), the Kemeny's rule (selection is made based on a ranking which minimizes the cumulative Hamming distance to the individuals' preference orders, see Kemeny (1959)), the Slater's rule (selection is made based on a ranking which minimizes the cumulative Hamming distance to the pairwise majority comparison results, see Slater (1961)), the Banks rule (The Banks winners are the top elements of all maximal transitive subtournaments with respect to inclusion, see Banks (1985) and Brandt et al. (2016)), and the Schwartz’s tournament equilibrium set rule (based on a recursive definition, see at length in Schwartz (1990)) are computationally intractable in worst cases since determining their respective winners is a NP-hard problem (Bartholdi et al. 1989a; Brandt et al. 2010; Conitzer 2006; Woeginger 2003). In contrast, the winner set of the proposed DSR-scoring method can be computed in polynomial time. According to ``\textit{if computing a solution is intractable, the applicability of the corresponding solution concept is seriously undermined}''(Brandt et al. 2010), therefore, \textbf{the proposed DSR-scoring method is superior to the Dodgson's rule, the Kemeny's rule, the Slater's rule, the Banks rule, and the Schwartz’s tournament equilibrium set (TEQ) rule in the narrow sense of computational complexity}.

\subsection{Relationships to the uncovered set, the top cycle set, the Smith set and the Schwartz set}

The uncovered set was independently proposed by Fishburn (1977) and Miller (1980). The term `uncovered set' in this subsection refers to Miller's uncovered set (Miller, 1980). Miller's uncovered set is defined by means of the covering relation over a tournament on finite alternative set $X$. Alternative $x$ \textit{covers} alternative $y$ iff every alternative less preferred to $y$ is also less preferred to $x$ (Miller, 1980). Formally, $x$ covers $y$ iff (Moulin, 1986, P275)
$$x\neq y, x\succ y, \mbox{~and~} \forall z\in X: y\succ z\rightarrow x\succ z.\eqno(11)$$
An uncovered set is composed of all alternatives that are not covered, that is, it is made up of  ``\textit{those points from which every other point is reachable in no more than two steps}'' (Miller, 1980, P74). 

As noted in Section 3.1, the requirements [$A\succeq B_1$, $\beta(A,B_1)>0$, $B_1\succ B_2$, $\beta(A,B_2)>0$] for tripartition  $\langle A,B_1,B_2\rangle$ in a tournament setting indicate that, there exists at least one alternative $x$ in $A$ such that $x$ covers every alternative in $B_1$. It may have occurred to the reader that there might be some relationship between the outcome of the DSR-scoring method and Miller's uncovered set. Surely in this case, and we may go further:

\textbf{Proposition 5} In a setting of tournament on finite alternative set $X$, $\forall x,y\in X, x\neq y: (x \mbox{~covers~} y)\rightarrow (\psi(x)>\psi(y))$, where $\psi(x)$ stands for the DSR-score of alternative $x$.

\textbf{Proof} By expression (6), we know that $\psi(x)=\sum_{w\in X}score_w(x)$ and $\psi(y)=\sum_{w\in X}score_w(y)$. If we have $\forall w\in X: score_w(x)\geq score_w(y)$, and $\exists w\in X: score_w(x)> score_w(y)$, then we will obtain the expected outcome $\psi(x)>\psi(y)$. We will achieve this goal by an exhaustive analysis of various aspects of possible partitions. We emphasize that the discussion is conducted in a tournament setting (i.e., connected and no ties).
\begin{itemize}
\item[(1)] Given expression (11), the alternative set can always be divisible with respect to $y$ and we will obtain $score_y(x)> score_y(y)$, since:
\begin{itemize}
\item[(a)] if $X_y^{\prec}=\emptyset$, by Def.4, then $\langle X_y^{\succ},\{y\}\rangle_y$ arises as a $y$-bipartition, where $X_y^{\succ}$ contains $x$ since we are given $x\succ y$. Because $x$ is in the top block and $y$ in the bottom block, by expressions (6-7) we know that $score_y(x)> score_y(y)$. Exactly, $score_y(x)=1$ and $score_y(y)=0$ in this situation.
\item[(b)] else (i.e., $X_y^{\prec}\neq\emptyset$),  first we know that $x\in X_y^{\succ}$ since we are given $x\succ y$. Further, we know that $\{x\}\succ X_y^{\prec}$ since $\{y\}\succ X_y^{\prec}$, $x\succ y$ and $x$ covers $y$. Noting that $x\in X_y^{\succ}$, by Def.4, we know that $\langle X_y^{\succ},\{y\},X_y^{\prec}\rangle_y$ is a $y$-tripartition. By expressions (6-7) we know that $score_y(x)> score_y(y)$. Exactly, $score_y(x)=\mu(x,y)+\sum_{w\in X_y^{\prec}}\mu(x,w)=1+|X_y^{\prec}|$ and $score_y(y)=\sum_{w\in X_y^{\prec}}\mu(y,w)=|X_y^{\prec}|$ in this situation.
\end{itemize}
\item[(2)] Regarding $x$, we will obtain $score_x(x)\geq score_x(y)$, since:
\begin{itemize}
\item[(a)] if $X$ is indivisible with respect to $x$, then under our scoring scheme (subsection 4.1) or by expressions (6-7) we know that $score_x(x)= score_x(y)=0$.
\item[(b)] if $X$ is divisible with respect to $x$, then $y$ must be included in the bottom block, whereas $x$ will be included in a top or middle block since we are given $x\succ y$. Under our scoring scheme (subsection 4.1) or by expressions (6-7) we know that $score_x(x)>0 = score_x(y)$.
\end{itemize}
\item[(3)] For $z\in X$, $z\neq x$ and $z\neq y$, we will obtain $score_z(x)\geq score_z(y)$, since:
\begin{itemize}
\item[(a)] if $X$ is indivisible with respect to $z$, then under our scoring scheme (subsection 4.1) or by expressions (6-7) we know that $score_z(x)= score_z(y)=0$.
\item[(b)] if $X$ is divisible with respect to $z$, then
\begin{itemize}
\item[(i)] if $X_z^{\prec}=\emptyset$, then the $z$-partition will be a bipartition, and both $x$ and $y$ are in the top block since in a tournament setting if $X_z^{\prec}=\emptyset$ then we have $x\succ z$ and $y\succ z$. Under our scoring scheme (subsection 4.1) or by expressions (6-7) we know that $score_z(x)= score_z(y)=1$ in this situation.
\item[(ii)] else 
\begin{itemize}
\item[--] if $y$ is in the bottom block of the $z$-partition, then under our scoring scheme (subsection 4.1) or by expressions (6-7) we know that $score_z(y)=0$. Because the value 0 is a minimal value that $score_a(b)$ possibly take at any alternative $a$ and any alternative $b$. Therefore, in this situation we obtain $score_z(x)\geq 0=score_z(y)$.
\item[--] otherwise, $y$ will be in the top block of the $z$-partition, since only $z$ can be in a possible middle block in the $z$-partition. Because the discussion is in a tournament setting (without ties), we thus know that, when $y$ is in the top block of $z$-partition, we must have $y\succ z$. Further, by expression (11), that is, since we are given ``$x$ covers $y$'', we thus have $x\succ z$, and $\forall w\in X_z^{\prec}: y\succ w\rightarrow x\succ w$. Under our scoring scheme (subsection 4.1) or by expressions (6-7) we know that $score_z(x)=\mu(x,z)+\sum_{w\in X_z^{\prec}}\mu(x,w)\geq\mu(y,z)+\sum_{w\in X_z^{\prec}}\mu(y,w)= score_z(y)$ in this situation.
\end{itemize}
\end{itemize}
\end{itemize}
\end{itemize}
Combining the results of the above steps (1)-(3),  as a consequence, we conclude that
$$score_y(x)+\sum_{w\in X\setminus\{y\}}score_w(x)> score_y(y)+\sum_{w\in X\setminus\{y\}}score_w(y)$$
and thus
$$\psi(x)=\sum_{w\in X}score_w(x)>\sum_{w\in X}score_w(y)=\psi(y),$$
so that the proposition is established. Q.E.D.

\vspace{0.2cm}

By virtue of Proposition 5 we have the following result.

\textbf{Proposition 6} In a finite tournament setting, the DSR-scoring method yields a choice set that is contained in the uncovered set. In other words, the winner of the DSR-scoring method belongs to the uncovered set. Notationally, suppose that $T(X)$ is a tournament on $X$. Let $UC(T(X))$ and $DSR(T(X))$ denote the solutions of the uncovered set and the solution of the DSR-scoring method, respectively. We have
$$DSR(T(X))\subseteq UC(T(X)).$$

\textbf{Proof}\footnote{The same line of reasoning was already used by Moulin when asserting ``\textit{A Copeland winner belongs to the uncovered set}'' (Moulin, 1986, P280).} From Proposition 5 we know that, for $x\neq y$,
$$(x \mbox{~covers~} y)\rightarrow (\psi(x)>\psi(y)),$$
which is equivalent to
$$\neg(\psi(x)>\psi(y))\rightarrow \neg(x \mbox{~covers~} y),$$
namely,
$$\mbox{if~}(\psi(y)\geq\psi(x))\mbox{~then~} (y \mbox{~is not covered by~} x).\eqno(*1)$$

The above statement (*1) holds for every $x$ provided $x\neq y$, namely,
$$\mbox{if~}[\psi(y)\geq\max_{x\in X\setminus\{y\}}\psi(x)]\mbox{~then~} [\forall x\in X\setminus\{y\}: (y \mbox{~is not covered by~} x)].\eqno(*2)$$

The above statement (*2) is a logical implication relation, where the antecedent [$\psi(y)\geq\max_{x\in X\setminus\{y\}}\psi(x)$] indicates that $y$ is a winner produced by the DSR-scoring method, and the consequent [$\forall x\in X\setminus\{y\}: (y \mbox{~is not covered by~} x)$] indicates that $y$ is an element of the uncovered set. Therefore, we know:
\begin{itemize}
\item the winner of the DSR-scoring method belongs to the uncovered set; namely,
\item $DSR(T(X))\subseteq UC(T(X))$,
\end{itemize}
so that the proposition is established. Q.E.D.

\vspace{0.2cm}

As an illustration, we reconsider Example 3, which is taken from Dutta (1988). As shown in subsection 5.2, the DSR-scoring method yields a choice set as $DSR(T(X))=\{a_1,a_2,a_3\}$. If we use the uncovered set method, as reported in Dutta (1988, P67), the uncovered set is $UC(T(X))=\{a_1,a_2,a_3,a_4,a_5,a_6\}$. Obviously, we have $DSR(T(X))\subset UC(T(X))$.

\vspace{0.2cm}

We now consider the relationships between the DSR-scoring winner set and several attractive choice set concepts (such as the top cycle set (Schwartz, 1972,1990) and the Smith set (Smith, 1973; Fishburn, 1977; Brandt, 2009) as well as the Schwartz set (Schwartz, 1972)), which are all defined based on dominance relation. To make the problem more manageable, we confine our discussion to the tournament settings. It is shown by Theorem 4 that the DSR-scoring method has the property of strong Gehrlein-stability. Since the strong Gehrlein-stability implies the Smith set principle, thus the DSR-scoring winner set is always included in the Smith set. In tournaments, the Smith set and the Schwartz set both coincide with the top cycle set (Brandt et al., 2007; Johnson, 2005). Hence the DSR-scoring winner set is also included both in the Schwartz set and the top cycle set in tournaments. 

However, as pointed out by Deb (1977), the top cycle set may contain Pareto inferior alternatives. In contrast, the DSR-scoring method satisfies both the weak Pareto property and the strong Pareto property. Therefore, \textbf{in tournaments the DSR-scoring winner set can be an interesting solution concept refining the top cycle and the Smith set as well as the Schwartz set}.

\subsection{Comparison with the Copeland method}

The Copeland method belongs to the Condorcet extension. Generally, it has several versions (Zwicker, 2016, P28), for example, $\mbox{Copeland}^0$, $\mbox{Copeland}^{\frac{1}{2}}$ and $\mbox{Copeland}^{\alpha}$ with $\alpha$ being a rational number in $[0,1]$. The first two mentioned are due to Copeland (1951), the last one is due to Faliszewski et al. (2007). They can be jointly interpreted in term of the parameter $\alpha$ as follows (Rothe and Schend, 2003). An alternative's Copeland score is assigned by means of pairwise majority comparisons with all the other alternatives: in a single pairwise contest with another alternative it will get one point whenever it wins and $\alpha$ point whenever they tie. In particular, $\mbox{Copeland}^0$ refers to the case where $\alpha=0$, and $\mbox{Copeland}^{\frac{1}{2}}$ the case where $\alpha=\frac{1}{2}$. In addition, there are still some other scoring schemes such as (1) in a pairwise contest the winner will get 1 point, while the loser will get $-1$ point (see, e.g, Klamler, 2005; Nurmi, 2012); and (2) in a pairwise contest the winner will get 2 point, while the loser will get $0$ point, and they both get $1$ point if they tie (Henriet, 1985).

The scoring formula in the DSR-scoring method for a single comparison (i.e., expression (5)) is similar to the scheme of Copeland$^{0.5}$. As remarked in Section 5, the properties stated in Lemma 1, Theorems 3 and 4 will remain unchanged when the DSR-scoring method is equipped with a $\alpha\in [0,1]$ as the scores assigned to the pair of contestants forming a tie. We denote this sort of variants of the DSR-scoring method by DSR$^\alpha$.

The DSR-scoring method and the Copeland method share many similar properties:
\begin{itemize}
\item[(1)] They both trivially satisfy the properties of anonymity, neutrality, and non-dictatorship. 
\item[(2)] They both satisfy the property of Pareto (as respectively shown by statements 1-2 of Theorem 3 and Gaertner (2009, P111)), the Condorcet winner principle (as respectively shown by statement 1 of Theorem 4 and Gaertner (2009, P111)), the Condorcet loser principle (as respectively shown by statement 2 of Theorem 4 and Felsenthal (2011)), the property of IIA within restricted domains (as respectively shown by statement 4 of Theorem 3 and Henriet (1985)), the strong Gehrlein-stability (as respectively shown by statement 3 of Theorem 4 and Barber$\grave{a}$ and Coelho (2008)).
\item[(3)] Both their winner sets are contained in the uncovered set (as respectively shown by Proposition 6 and Moulin (1986, P280)).
\end{itemize}

The main differences between the DSR-scoring method and the Copeland method include, but not limit to
\begin{itemize}
\item[(1)] They differ in the computational complexity. Particularly, the Copeland method runs in polynomial time $O(m^2)$ (Brandt et al., 2016; Hudry, 2009), whereas the DSR-scoring method runs in polynomial time $O(m^3)$ (Corollary 1), where $m$ represents the number of alternatives. 
\item[(2)] They may yield very different outcomes when applied to the same problem. We demonstrate this on an example which is taken from Brandt et al. (2016). 

\textbf{Example 5} Consider a tournament on the alternative set $X=\{a,b,c,d\}$. The dominance relation and the tournament matrix are included in Figure 4, where `$x\rightarrow y$' indicates `$x\succ y$'. If we use the DSR-scoring method, the analysis process is shown by Table 7.

\begin{figure}[H]
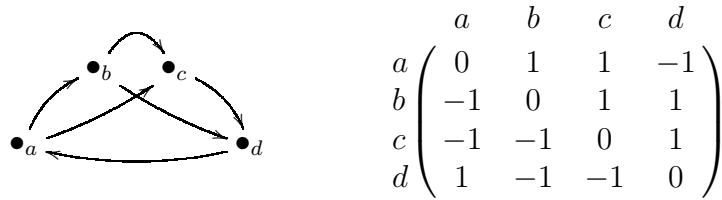

\centering
\subfigure{}
{
\begin{minipage}{5cm}
\[ \xygraph{  
        !{<0cm,0cm>;<1cm,0cm>:<0cm,1cm>::}  
        !{(0,0) }*+{\bullet_{a}}="a"  
        !{(1,1) }*+{\bullet_{b}}="b"  
        !{(2,1) }*+{\bullet_{c}}="c"  
        !{(3,0)}*+{\bullet_{d}}="d"  
            "a":@/^0.2cm/"b"^(0.6){}  
            "a":@/_0.1cm/"c"^(0.6){}
            "b":@/_0.1cm/"d" ^(0.8){}   
            "c":@/^0.2cm/"d" ^(0.8){}  
            "d":@/^/"a" ^(0.4){}  
            "b":@/^0.5cm/"c" ^(0.7){}      
        }
\]  
\end{minipage}
}
\subfigure{}
{
\begin{minipage}{5cm}
\[
\bordermatrix{
    & a & b & c & d\cr
  a & 0 & 1 & 1 & -1\cr
  b & -1 & 0 & 1 & 1\cr
  c & -1 & -1 & 0 & 1\cr
  d & 1 & -1 & -1 & 0
}
\]
\end{minipage}
}
\caption{Tournament diagram and tournament matrix in Example 5.}
\end{figure}

\begin{table}[h]
\renewcommand\arraystretch{1.3}
\centering\footnotesize
\caption{Analysis process of Example 5}
\begin{tabular}{ccccccc}
    \toprule
    \multicolumn{1}{c}{$a\succ b, a\succ c, b\succ c,$}  & \multicolumn{4}{c}{alternative ($x$)} \\
    \cmidrule{2-3}  \cmidrule{4-5}  \cmidrule{6-7}
    {$b\succ d, c\succ d, d\succ a$}  & $a$ & $b$ & $c$ & $d$ \\
    \midrule
$X_x^{\succeq,\neq}$ &

$\{d\}
$ &
$\{a\}
$& $\{a,b\}$ & $\{b,c\}$ \\

$\{x\}$ &

$\{a\}
$ &
$\{b\}
$& $\{c\}$ & $\{d\}$ \\

$X_a^{\prec}$ &

$\{b,c\}
$ &
$\{c,d\}
$& $\{d\}$ & $\{a\}$ \\

$x$-partition &

{-} & {$\left \langle {\begin{array}{cc} {\{a,b\}}\\
{\{c,d\}}
\end{array}}\right \rangle$$_b$} & {$\left \langle {\begin{array}{cc} {\{a,b\}}\\
{\{c\}}\\
{\{d\}}
\end{array}}\right \rangle$$_c$} & {-} \\

$score_x(a)$ &

{0} & {2} & {1} & {0} \\

$score_x(b)$ &

{0} & {2} & {2} & {0} \\

$score_x(c)$ &

{0} & {0} & {1} & {0} \\

$score_x(d)$ &

{0} & {0} & {0} & {0} \\

    \bottomrule
\end{tabular}
\end{table}

By expression (6), we obtain the DSR-scores of alternatives:
$$\begin{array}{cc}
\psi(a)=\sum_{x\in \{a,b,c,d\}}score_x(a)=3,& \psi(b)=\sum_{x\in \{a,b,c,d\}}score_x(b)=4,\\
\psi(c)=\sum_{x\in \{a,b,c,d\}}score_x(c)=1,& \psi(d)=\sum_{x\in \{a,b,c,d\}}score_x(d)=0,
\end{array}$$
so that the tournament solution is $\{b\}$ by using the DSR-scoring method. 

In contrast, if we apply some conventional methods to Example 5, such as the Copeland method, the Slater method, the Markov chain method and the uncovered set method, as reported in Brandt et al. (2016), the tournament solutions will be $\{a,b\}$, $\{a\}$, $\{a\}$ and $\{a,b,d\}$, respectively.
\end{itemize}

The above comparison and analysis show that, the proposed DSR-scoring method shares almost every feature with the Copeland method, except that the DSR-scoring method is of a higher computational complexity. Even though the outcomes produced by the two methods may be quite different (as illustrated Example 5), a question which arises is, whether the DSR-scoring method still has some usefulness?  The answer might be affirmative. From another perspective, complexity may have some significance. This paper does not consider strategic play. However, as it is well known, manipulation and strategy-proofness is a very appealing topic in many subjects such as social choice and mechanism design (Moulin, 1980; Maskin, 2008). It is asserted by the famous Gibbard-Satterthwaite theorem (Gibbard, 1973; Satterthwaite, 1975), that any single-valued social choice function (SCF, a mapping from individual preferences to outcomes), which is defined on unrestricted domain of preference orderings and whose range contains at least three outcomes, is either dictatorial or manipulable\footnote{A SCF is called manipulable if an individual (or a coalition of individuals) can benefit from misreporting her (or their) preference (or preferences).}. Although any single-valued SCF is susceptible to manipulation in general, however, different SCFs can have different levels of difficulties to be manipulated (Conitzer and Sandholm, 2005; Xia and Conitzer, 2008). Indeed, if it is difficult to determine a winner for a SCF, it is not easy to manipulate it either. It is natural to use the computational complexity as a shield against manipulation, as shown by Bartholdi et al. (1989b). \textbf{In the narrow sense of strategy-proofness, therefore, the proposed DSR-scoring method is superior to the Copeland method, since a higher computational complexity is required to determine the winner set of the DSR-scoring method}.

In addition, as shown by Example 5, the winner set of the DSR-scoring method can be contained in the winner set of the Copeland method in some cases. Indeed, we have the following result.

\textbf{Proposition 7} In a tournament where the number of alternatives is not more than 4, the DSR-scoring method yields a winner set that is contained in the winner set of the Copeland method.

\textbf{Proof} The argument holds in a trivial way for two-alternative cases or else, for three-alternative cases, it can be easily derived from Lemma 1'. If there are 4 alternatives in the tournament, we use proof by contradiction, and so we need to show that for any alternative $y$, if it is not a Copeland winner then it is not a DSR-scoring winner either. We distinguish two cases.
\begin{itemize}
\item[(1)] If $y$ is not included in the uncovered set, then it is not a winner of the DSR-scoring method, since Proposition 6 asserts that a winner of the DSR-scoring method should be in the uncovered set. As a consequence, in this case the argument holds.
\item[(2)] Else, our assumption will be the case that $y$ is not a Copeland winner but it is included in the uncovered set. Let $x$, different from $y$, denote a Copeland winner. Now we need to show that the DSR-score of $x$ is greater than the DSR-score of $y$, that is, we need to show $\psi(x)>\psi(y)$ under the given conditions.

According to Moulin (1986, P280), we know that a Copeland winner is always included in the uncovered set, so that $x$ is in the uncovered set. Assume that the tournament is composed of $x$, $y$, $z$ and $u$.
\begin{itemize}
\item On one hand, we assume $x\succ y$. In this situation, we must exclude the case that $y$ beats none, since if $y$ beats none then $y$ will be covered by all the other alternatives in a tournament setting. However, we are given that $y$ is included in the uncovered set. Hence we know that $X_y^{\prec}\neq\emptyset$ in the sense that there ought to be some alternatives that are beaten by $y$. We can assume $y\succ z$ and further examine what kinds of dominance relationships ought to be between the pair $x$ and $z$, the pair $x$ and $u$, and the pair $y$ and $u$, as well as the pair $z$ and $u$. First, we will have $u\succ y$, since otherwise $x$ would cover $y$ in order to become a Copeland winner given that $y$ is not a Copeland winner (particularly, if $y\succ u$ then the Copeland score of $y$ would be 2, while $x$ would have to have a Copeland score value of 3 to exceed $y$). But we are given that $y$ is in the uncovered set, thus we will have $u\succ y$. Now we know $X_y^{\prec}=\{z\}$. Second, we further have $z\succ x$, since otherwise $x$ would cover $y$ given $x\succ y$, $y\succ z$ and $X_y^{\prec}=\{z\}$. Furthermore, we must have $z\succ u$, since otherwise $u$ would cover $y$ given $u\succ y$, $y\succ z$ and $X_y^{\prec}=\{z\}$. Finally, we must have $x\succ u$ for $x$ to become a Copeland winner so that $x$ has a greater Copeland score than $y$.

The above discussion leads to a unique tournament on $\{x,y,z,u\}$, where $x\succ y$, $x\succ u$, $z\succ x$, $y\succ z$, $u\succ y$, and $z\succ u$, as depicted in Figure 5. Clearly, if we rename $x$ by $b$, $y$ by $d$, $z$ by $a$, and $u$ by $c$, then the derived tournament is none other than the tournament in Example 5. According to the analysis process shown by Table 7, we know $\psi(x)=4>0=\psi(y)$, so that the argument holds in the case of $x\succ y$.
\begin{figure}[H]
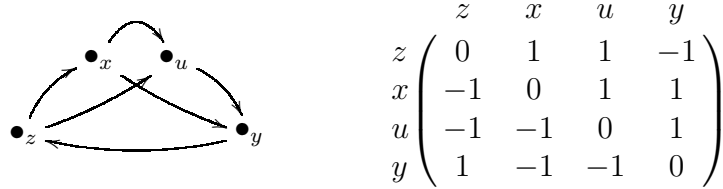

\centering
\subfigure{}
{
\begin{minipage}{5cm}
\[ \xygraph{  
        !{<0cm,0cm>;<1cm,0cm>:<0cm,1cm>::}  
        !{(0,0) }*+{\bullet_{z}}="z"  
        !{(1,1) }*+{\bullet_{x}}="x"  
        !{(2,1) }*+{\bullet_{u}}="u"  
        !{(3,0)}*+{\bullet_{y}}="y"  
            "z":@/^0.2cm/"x"^(0.6){}  
            "z":@/_0.1cm/"u"^(0.6){}
            "x":@/_0.1cm/"y" ^(0.8){}   
            "u":@/^0.2cm/"y" ^(0.8){}  
            "y":@/^/"z" ^(0.4){}  
            "x":@/^0.5cm/"u" ^(0.7){}      
        }
\]  
\end{minipage}
}
\subfigure{}
{
\begin{minipage}{5cm}
\[
\bordermatrix{
    & z & x & u & y\cr
  z & 0 & 1 & 1 & -1\cr
  x & -1 & 0 & 1 & 1\cr
  u & -1 & -1 & 0 & 1\cr
  y & 1 & -1 & -1 & 0
}
\]
\end{minipage}
}
\caption{The tournament for the case $x\succ y$ in the proof of Proposition 7.}
\end{figure}

\item On the other hand, we analyze the case of $y\succ x$ given that both $x$ and $y$ are included in the uncovered set, and $x$ is a Copeland winner while $y$ is not. Given $y\succ x$, for $x$ to exceed $y$ with regard to Copeland score, we must have $|X_x^{\prec}|>1$ in the sense that there ought to be at least two alternatives that are beaten by $x$. As the tournament is defined on $\{x,y,z,u\}$ and we are given $y\succ x$, we thus know that $X_x^{\prec}=\{z,u\}$, which is interpreted as $x\succ z$ and $x\succ u$. Next we need to determine what kinds of dominance relationships ought to be between the pair $y$ and $z$, the pair $y$ and $u$, and the pair $z$ and $u$. Since $x$ should have a greater Copeland score than $y$, we thus know that $y$ should not dominate any of $z$ and $u$, so that we have $z\succ y$ and $u\succ y$. Regarding the dominance relationship between the pair $z$ and $u$, we can suppose either $z\succ u$ or $u\succ z$ since none of these assumptions will cause conflicts with known conditions. As a result, we obtain two tournaments as depicted in Figure 6.
\begin{figure}[H]
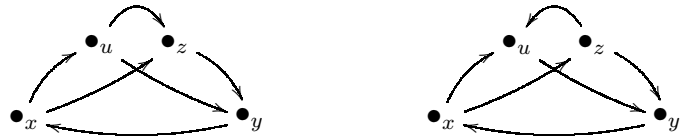

\centering
\subfigure{}
{
\begin{minipage}{5cm}
\[ \xygraph{  
        !{<0cm,0cm>;<1cm,0cm>:<0cm,1cm>::}  
        !{(0,0) }*+{\bullet_{x}}="x"  
        !{(1,1) }*+{\bullet_{u}}="u"  
        !{(2,1) }*+{\bullet_{z}}="z"  
        !{(3,0)}*+{\bullet_{y}}="y"  
            "x":@/^0.2cm/"u"^(0.6){}  
            "x":@/_0.1cm/"z"^(0.6){}
            "u":@/_0.1cm/"y" ^(0.8){}   
            "z":@/^0.2cm/"y" ^(0.8){}  
            "y":@/^/"x" ^(0.4){}  
            "u":@/^0.5cm/"z" ^(0.7){}      
        }
\]
\end{minipage}
}
\subfigure{}
{
\begin{minipage}{5cm}
\[ \xygraph{  
        !{<0cm,0cm>;<1cm,0cm>:<0cm,1cm>::}  
        !{(0,0) }*+{\bullet_{x}}="x"  
        !{(1,1) }*+{\bullet_{u}}="u"  
        !{(2,1) }*+{\bullet_{z}}="z"  
        !{(3,0)}*+{\bullet_{y}}="y"  
            "x":@/^0.2cm/"u"^(0.6){}  
            "x":@/_0.1cm/"z"^(0.6){}
            "u":@/_0.1cm/"y" ^(0.8){}   
            "z":@/^0.2cm/"y" ^(0.8){}  
            "y":@/^/"x" ^(0.4){}  
            "z":@/_0.5cm/"u" ^(0.7){}      
        }
\]
\end{minipage}
}
\caption{The tournaments for the case $y\succ x$ in the proof of Proposition 7.}
\end{figure}
One can see that the left picture in Figure 6 is exactly the same as that in Figure 4 from a structural perspective. Following a same analytic process shown in Table 7, we have $\psi(x)=3>0=\psi(y)$, so that the argument holds in the case of $y\succ x$ and $u\succ z$. With respect to the right picture in Figure 6, through a similar procedure we obtain $\psi(x)=3>0=\psi(y)$, so that the argument holds in the case of $y\succ x$ and $z\succ u$. Therefore, the argument holds in the case of $y\succ x$ when either given $u\succ z$ or $z\succ u$.
\end{itemize}
We conclude that the argument holds when either $x\succ y$ or $y\succ x$ provided that $y$ is included in the uncovered set.
\end{itemize}
We conclude that the argument holds no matter whether $y$ is included in the uncovered set or not, so that the proposition is established. Q.E.D.

\vspace{0.2cm}

Here it should be noted that, different $\alpha$ may make the DSR-scoring method produce different orderings. We demonstrate this by a simple example.

\textbf{Example 6} Consider a complete preference relation over $X=\{a,b,c\}$, as shown by Figure 7. We denote by $\alpha$ the point to be assigned to the tied pair. We use the DSR-scoring method and summarize the procedures into Table 8.

\begin{figure}[H]
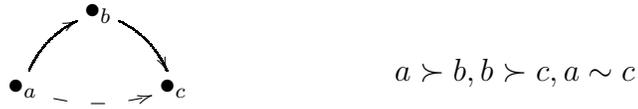

\centering
\subfigure{}
{
\begin{minipage}{5cm}
\[ \xygraph{  
        !{<0cm,0cm>;<1cm,0cm>:<0cm,1cm>::}  
        !{(0,0) }*+{\bullet_{a}}="a"  
        !{(1,1) }*+{\bullet_{b}}="b"  
        !{(2,0) }*+{\bullet_{c}}="c"    
            "a":@/^0.2cm/"b"^(0.4){}
            "b":@/^0.2cm/"c"^(0.4){}
            "a":@/_0.2cm/@{-->}"c" ^(0.4){}   
        }
\]  
\end{minipage}
}
\subfigure{}
{
\begin{minipage}{5cm}
\[
\begin{array}{ccc}
a\succ b, b\succ c, a\sim c
\end{array}
\]
\end{minipage}
}
\caption{Preference relation of Example 6.}
\end{figure}

\begin{table}[h]
\renewcommand\arraystretch{1.3}
\centering\footnotesize
\caption{Computing $score_x(y)$ in Example 6}
\begin{tabular}{ccccccc}
    \toprule
    \multicolumn{1}{c}{Preference relation}  & \multicolumn{3}{c}{alternative ($x$)} &\\
    \cmidrule{2-3}  \cmidrule{4-5} \cmidrule{6-7}
    {$a\succ b, b\succ c, a\sim c$}  & $a$ & $b$ & $c$ &\\
    \midrule
$X_x^{\succeq,\neq}$ &

$\{c\}
$ &
$\{a\}
$& $\{a,b\}$ &\\

$\{x\}$ &

$\{a\}
$ &
$\{b\}
$& $\{c\}$ &\\

$X_x^{\prec}$ &

$\{b\}
$ &
$\{c\}
$& $\{\}$ &\\

$x$-partition &

{--} & {$\left \langle {\begin{array}{cc} {\{a,b\}}&\\
{\{c\}}
\end{array}}\right \rangle$$_b$} & {$\left \langle {\begin{array}{cc} {\{a,b\}}&\\
{\{c\}}
\end{array}}\right \rangle$$_c$} &\\

$score_x(a)$ &

{0} & {1+$\alpha$} & {$\alpha$} \\

$score_x(b)$ &

{0} & {1} & {1} \\

$score_x(c)$ &

{0} & {0} & {0} \\

    \bottomrule
\end{tabular}
\end{table}

By expression (6), we obtain the DSR-scores of alternatives:
$$
\psi(a)=\sum_{x\in \{a,b,c\}}score_x(a)=1+2\alpha, \psi(b)=\sum_{x\in \{a,b,c\}}score_x(b)=2, \psi(c)=\sum_{x\in \{a,b,c\}}score_x(c)=0.
$$

Clearly, the final ranking depends on the value of $\alpha$:
\[
\begin{cases}
0\leq\alpha<0.5: & b\succ a\succ c,\\
\alpha=0.5: & a\sim b\succ c, \\
0.5<\alpha\leq 1: & a\succ b\succ c.
\end{cases}
\]

Similar phenomenon can also be observed when applying the Copeland$^\alpha$ method\footnote{As point out by Moulin(1986, P273), ``\textit{the Copeland score can be defined in several nonequivalent ways (by counting one half point or zero point in case of a tie)}''.} to the data in Example 6, since the alternatives' Copeland scores are as follows.
\[
\left. \begin{array}{lll} \mbox{Copeland score of~} a: & 1+\alpha,\\ \mbox{Copeland score of~} b: & 1, \\ \mbox{Copeland score of~} c: & \alpha.\end{array} \right\}\Rightarrow
\begin{cases}
\alpha=0: & a\sim b\succ c,\\
0<\alpha<1: & a\succ b\succ c, \\
\alpha=1: & a\succ b\sim c.
\end{cases}
\]

Therefore, for practical use, how to determine the value of parameter $\alpha$ deserves further consideration. The calibration of $\alpha$ may be performed by virtue of a popular `benchmark' (depending on the application scenario). For instance, suppose that the data in Example 6 are collected from football contests among three teams $a$, $b$ and $c$. Currently, a customary international practice is conducted as follows: for a pair of contestant, the winner will get 3 points, and the loser will get 0 points; if the game ends in a draw, both sides  will receive 1 point each. By virtue of this convention, the ranking of the three teams will be $a\succ b\succ c$. Now we turn to the DSR-scoring method. Clearly, if $\alpha$ takes a value from $(0.5, 1)$ then the alternatives' final ranking will be $a\succ b\succ c$ as expected. Therefore, in an application scenario of football contest, $\alpha$ can take a value from $(0.5,1)$. For example, 0.75 can be an interesting possibility.

\section{Conclusion}

This paper introduced several ordered partitions of the alternative set (called dominating-set-relaxed partitions, for short, DSRPs) by relaxing the requirements of dominating set. By using the DSRPs, a method (called the DSR-scoring method) was designed for deriving an overall ranking of the alternatives or a choice set from a complete binary preference relation on a finite alternative set. Here the complete binary preference relation, i.e., the input data of the DSR-scoring method, can be a pairwise majority relation or a tournament. It has some formally attractive properties and possesses some competitive advantages which can be summarized as follows.
\begin{itemize}
\item [(1)] The proposed DSR-scoring method satisfies a number of desirable properties including Condorcet winner principle, Condorcet loser principle, strong Gehrlein-stability (hence Smith set principle), anonymity, neutrality, weak Pareto, strong Pareto, non-dictatorship, and [independence of irrelevant alternatives (IIA) when the pairwise majority relation is an ordering on the alternative set].

\item [(2)] The proposed DSR-scoring method is superior to some well-studied voting rules (such as the Dodgson's rule, the Kemeny's rule, the Slater's rule, the Banks rule, and the Schwartz’s tournament equilibrium set (TEQ) rule) in the narrow sense of computational complexity, since determining their respective winners of these well-known rules is a NP-hard problem, while the winner set of the DSR-scoring method can be computed in polynomial time. 

\item [(3)] If the input of the DSR-scoring method is a complete and transitive majority relation, then the collective preference relation determined by the DSR-scoring method coincides with that of the input majority relation. And thus the proposed DSR-scoring method shares the same collective preference function on the dichotomous domain with the approval voting and the majority voting.

\item [(4)] When the proposed DSR-scoring method is used for deriving a choice set from a tournament, its winner belongs to the uncovered set, the top cycle set, the Smith set, and the Schwartz set. In addition, in a tournament where the number of alternatives is not more than 4, the DSR-scoring method yields a winner set that lies inside the winner set of the Copeland method.

\item [(5)] In the narrow sense of strategy-proofness, the proposed DSR-scoring method is superior to the Copeland method, since a higher computational complexity is required to determine the winner set of the DSR-scoring method.

\end{itemize}

In addition to the above attractive properties, the proposed DSR-scoring method also has some disadvantages. For example, since the method belongs to the Condorcet extension family, and as pointed out by Moulin (1988) that ``\textit{Condorcet's principle implies the no-show paradox}'', therefore, the DSR-scoring method does not guarantee the exclusion of no-show paradox.

It is worth noting that, we have proved (Proposition 7) and illustrated (Example 5) that, in tournaments, the winner set of the proposed method is a subset, sometimes proper, of the Copeland winner set when the number of alternatives is not more than four. Thus the winner set of the proposed method offers a potential refinement of the Copeland winner set in no-more-than-four-alternative tournaments. We conjecture that similar results would hold for four-more-alternative tournaments (Indeed, one can check that the argument holds for five-alternative tournaments), but we leave the proof (Intuitively, the technique of mathematical induction will be an interesting possibility) to future work.

\vspace{0.1cm}


 
\section*{Competing interests}
The author declares that he has no competing interests.

\end{document}